\documentclass[pre,twocolumn]{revtex4}
\usepackage{amssymb,amsmath}
\usepackage{epsfig}
\usepackage{graphicx}
\usepackage{latexsym}

\newcommand{\bit}{\begin{itemize}}
\newcommand{\eit}{\end{itemize}}
\newcommand{\beq}{\begin{equation}}
\newcommand{\eeq}{\end{equation}}
\newcommand{\bea}{\begin{eqnarray}}
\newcommand{\eea}{\end{eqnarray}}

\begin{document}
\title{Strain localization in a shear transformation zone model for amorphous solids}
\author{M. L. Manning}%
 \email{mmanning@physics.ucsb.edu}
\author{J. S. Langer}
\author{J. M. Carlson}
\affiliation{%
Department of Physics, University of California, Santa Barbara, 93106
}%
\date{\today}

\begin{abstract}We model a sheared disordered solid using the theory of Shear Transformation Zones (STZs). In this mean-field continuum model the density of zones is governed by an effective temperature that approaches a steady state value as energy is dissipated.  We compare the STZ model to simulations by Shi, {\em et al.}[Y. Shi {\em et al.} PRL 98, 185505 (2007)], finding that the model generates solutions that fit the data, exhibit strain localization, and capture important features of the localization process. We show that perturbations to the effective temperature grow due to an instability in the transient dynamics, but unstable systems do not always develop shear bands. Nonlinear energy dissipation processes interact with perturbation growth to determine whether a material exhibits strain localization. By estimating the effects of these interactions, we derive a criterion that determines which materials exhibit shear bands based on the initial conditions alone. We also show that the shear band width is not set by an inherent diffusion length scale but instead by a dynamical scale that depends on the imposed strain rate.
\end{abstract}

\maketitle

\section{Introduction}
\label{Intro}

Amorphous materials often comprise or lubricate sheared material interfaces and require more complicated constitutive equations than simple fluids or crystalline solids.  They flow like a fluid under large stresses, creep or remain stationary under smaller stresses, and have complex, history-dependent behavior. Bulk metallic glasses, granular materials, and bubble rafts are just some of the disordered materials that exhibit these features~\cite{Lu, anandar, Lauridsen}.

  In this paper, we focus on {\em strain localization}, the spontaneous development of coexisting flowing and stationary regions in a sheared material. Strain localization has been identified and studied experimentally in granular materials~\cite{Tsai_Gollub,fenistein}, bubble rafts~\cite{Lauridsen,Kabla}, complex fluids~\cite{mair,micelle2}, and bulk metallic glasses~\cite{Lu,Johnson2}. Shear banding may play an important role in the failure modes of structural materials and earthquake faults.  Localization is a precursor to fracture in bulk metallic glasses~\cite{Lu} and has been cited as a mechanism for material weakening in granular fault gouge on faults~\cite{Marone}. 

We model sheared material interfaces using the theory of Shear Transformation Zones (STZs). Spaepen first postulated that plastic flow could be modeled by tracking localized zones or regions of disorder~\cite{Spaepen}. Argon and Bulatov developed mean-field equations for these localized regions~\cite{Argon, Bulatov}.  Building on this work, Falk and Langer~\cite{Falk_L1} introduced a mean-field theory which postulates that the zones, which correspond to particle configurations that are more susceptible to plastic rearrangements than the surrounding particles, could reversibly switch between two orientational states in response to stress.  Additionally, these zones are created and annihilated as energy is dissipated in the system. This theory captures many of the features seen in experiments, such as work-hardening and yield stress~\cite{Falk_L2,Pechenik, Bouchbinder}.

There are several complementary approaches to understanding disordered solids.  On the smallest scales, molecular dynamics and quasi-static simulations generate a wealth of information about particle interactions and emergent macroscopic behavior --- including shear banding~\cite{Xu_OHern2,Shi,Varnik} ---  but they are limited to smaller numbers of particles and time scales. On the largest scales, phenomenological models such as viscoelasticity and the Dieterich-Ruina friction law~\cite{Dieterich,Ruina}, which has been studied extensively in the context of rock mechanics, describe stress-strain step and frequency responses, but to date these laws have not been derived from microscopic dynamics.  

A third approach focuses on the dynamics of collections of particles and their configurations.   Examples of this approach include the Spot Model for granular materials~\cite{Bazant}, Soft Glassy Rheology (SGR)~\cite{Sollich}, and the theory of Shear Transformation Zones. The Spot Model generalizes dislocation dynamics in crystalline solids by postulating that a {\em spot}, a region of extra free volume shared among a collection of particles, executes a random walk through the material. Both SGR and STZ theories postulate that there are meso-scale {\em configurational soft spots} which are more susceptible than the surrounding material to yield under shear stress. 

  Recently, Langer postulated that an effective temperature should govern the density of shear transformation zones~\cite{Langer_Eff}. Mehta and Edwards~\cite{Edwards} were perhaps the first to point out that although thermal temperature does not determine statistical distributions for macroscopic particles such as powders, the statistical properties of these systems could still be characterized by a small number of macroscopic state variables, such as free volume.   Cugliandolo, Kurchan, and Peliti~\cite{Cugliandolo} show that heat flow in these slowly stirred systems is determined by an {\em effective} temperature. Several studies have confirmed that effective temperature can be used to characterize slowly sheared  granular packings and bubble rafts~\cite{Ono,OHern}.   

For many systems thermal temperature is not sufficient to cause configurational rearrangements, but slow shearing causes the particles to ergodically explore configuration space. Although we make no attempt to calculate the entropy from first principles, we postulate that because the material is being sheared slowly, configurational (as opposed to kinetic) degrees of freedom are the dominant contributions to the entropy. The statistical distribution of steady state configurations should maximize this entropy and be described by an effective temperature.

The effective temperature is an important macroscopic state variable in this STZ description. It governs the properties of statistical distributions for configurational degrees of freedom in the material, such as density fluctuations. Langer hypothesized that shear transformation zones are unlikely, high-energy density fluctuations, and therefore the number of such zones in a system should be proportional to a Boltzmann factor which here is a function of the effective temperature instead of the usual thermal temperature~\cite{Langer_Eff}.  In this manner, the density of shear transformation zones is related to the effective temperature and therefore the plastic flow is coupled to the local disorder. This is a mechanism for strain localization because a region with higher disorder is more susceptible to flow under stress and flowing regions become more disordered. 
  
  The effective temperature mechanism for strain localization is different from other proposed mechanisms governed by thermal temperature or free volume. Griggs and Baker~\cite{Griggs} proposed that the strong temperature dependence of viscosity near the glass transition leads to thermal softening and strain localization. In that formulation the heat generated by plastic work raises the thermal/kinetic temperature, instead of the configurational/effective temperature. Lewandowski and Greer~\cite{Lewandowski} have shown that in a bulk metallic glass the thermal temperature diffuses too quickly to control shear localization in an adiabatic description, though Braeck and Podladchikov~\cite{Braeck} have shown that a non-adiabatic thermal theory can explain the data. 

Alternatively, several authors have suggested that the free volume governs strain localization~\cite{Anael1, Johnson2}. In these descriptions, plastic deformation in the solid is accompanied by a larger free volume, which in turn softens the material. However, the exact relationship between the free volume and plastic deformation is difficult to determine. In this paper we suggest an alternative description: the heat generated by plastic deformation is dissipated in the configurational degrees of freedom, according to the first law of thermodynamics. The local free volume is {\em related} to the effective temperature (for example, in a Lennard-Jones glass the particles can become more disordered by moving away from their equilibrium separations), but the effective temperature is the relevant macroscopic state variable governing particle dynamics. 

 The STZ formulation in this paper is {\em athermal} --  the thermal temperature $T$ does not cause any configurational rearrangements. This is an approximation which is clearly appropriate for bubble rafts and granular materials and may be relevant for metallic glasses well below the glass transition temperature.  The athermal approximation significantly simplifies the STZ equations and clarifies the nature of the instability that leads to localization, but the resulting description does not capture features such as thermally activated creep or relaxation. As we discuss in Section~\ref{long_time}, some behaviors of real materials and simulations are likely due to thermal activation or relaxation and will not be accurately represented in this athermal STZ formulation.

  In this paper we show how athermal STZ theory with effective temperature generates solutions with strain localization. This paper is divided into sections as follows:  Section~\ref{STZeq} develops the constitutive equations for athermal STZ theory with effective temperature in a simple shear geometry.  Section~\ref{Numerical} compares numerical results from the continuum STZ theory to simulations of Lennard-Jones glasses by Shi, {\em et al.}  We compare macroscopic stress-strain curves for different initial conditions and compare the strain rate and potential energy as a function of position and time inside the sheared material.  Section~\ref{Stability} investigates the stability properties of the STZ equations, and presents a description of how shear bands form and remain intact for long times. We show that an instability in the transient dynamics amplifies perturbations to the effective temperature, and that this process interacts with energy dissipation processes to determine the long-time behavior of the strain rate field.  Section~\ref{Conclusions} concludes the paper with a discussion of our results and future directions.

\section{Review of STZ equations}
\label{STZeq}
 
A mean field theory for shear transformation zones, first derived by Argon~\cite{Argon}, has been developed in a series of papers~\cite{Falk_L1, Falk_L2,Langer_Eff, Bouchbinder}. For completeness, this section reviews the athermal STZ equations and emphasizes their similarity to other models for elastic and plastic flow. Section~\ref{General} introduces and motivates the STZ equations. Section~\ref{InfiniteStrip} tailors the equations to describe a particular geometry -- an infinite strip of material driven from the boundaries at a velocity $V_0$. 

\subsection{General equations}
\label{General}
When describing systems that undergo plastic and elastic deformation, it is convenient to break up the total stress tensor into hydrostatic and deviatoric components, $\sigma_{i j} = - p \delta_{i j} + \overline{\mu} s_{i j}$, because hydrostatic stress generally does not cause yield. To simplify notation, the deviatoric stress has been nondimensionalized by an effective shear modulus $\overline{\mu}$ that characterizes the stiffness of the STZs. In STZ theory, the stress scale $\overline{\mu}$ is important because it characterizes the stress at which the material begins to yield. In the slowly sheared materials we are modeling, the speed of sound in the material is very fast compared to the rate of plastic deformation. In this case the stress gradients equilibrate very quickly, and we take the zero density limit of the momentum conservation equations. This results in static elastic equations for the stress:

\beq
\label{stat_eq}
\frac{\partial \sigma_{i j}}{\partial x_{j}} = 0.
\eeq
We further assume that the rate of deformation tensor is the sum of elastic and plastic parts:
\bea
\label{Dtotal}
D^{total}_{i j} &=& \frac{1}{2} \left( \frac{\partial v_i}{\partial x_j} + 
 \frac{\partial v_j}{\partial x_i} \right)  \nonumber \\
 &=& \frac{{\cal D}}{{\cal D} t} \left( -\frac{p}{2 K} \delta_{i j} + \frac{\overline{ \mu}}{2 \mu} s_{i j} \right) + D^{plast}_{i j},
\eea
 where ${\cal D} / {\cal D} t $ is the material or corotational derivative defined as:
\begin{equation} 
\frac{{\cal D} A_{ij}} {{\cal D} t} = \frac{\partial A_{ij}}{ \partial t} + v_k \frac{\partial A_{ij}}{ \partial x_k}  + A_{ik}\omega_{kj} - \omega_{ik} A_{kj},
\end{equation}
and $\omega_{ij} = 1/2 ( \partial v_i / \partial x_j - \partial v_j / \partial x_i)$. We also use the notation $\dot A_{ij} = \partial A_{ij} / \partial t$. 

The plastic rate of deformation tensor can be written in terms of dynamical variables from STZ theory. We postulate that under shear stress, each STZ deforms to accommodate a certain amount of shear strain, and cannot deform further in the same direction. This is modeled by requiring that each STZ be in one of two states: oriented along the principle stress axis in the direction of applied shear, which we will denote ``$+$'', or in the perpendicular direction, ``$-$''.

 Under applied strain, the STZ will {\em flip} in the direction of strain, from ``$-$'' to ``$+$''. Under shear stress in the opposite direction, the STZs can revert to their original configurations, which corresponds to a flip from ``$+$'' to ``$-$''. These rearrangements or {\em flips} occur at a rate which depends on the stress $R(s)$ and a characteristic attempt frequency, $\tau_0$.  Because each STZ can flip at most once in the direction of applied strain, STZs must be created and annihilated to sustain plastic flow.   Based on these considerations, the number density of STZs in each direction, $n_{\pm}$ obeys the following differential equation
\beq
\label{number_density}
\tau_0 \dot{n}_{\pm} = R(\pm s) n_{\mp} - R(\mp s) n_{\pm} + \Gamma \left( \frac{n_{\infty}}{2} e^{-1/ \chi} - n_{\pm} \right), 
\eeq
where $R(\pm s)/ \tau_0$ is the rate of switching per STZ as a function of stress, $\Gamma$ is the rate at which energy is dissipated per STZ, and $n_{\infty} e^{-1/ \chi}$ is the steady state density of STZs in equilibrium. 

The first two terms in Eq.~(\ref{number_density}) correspond to STZs switching from ``$+$'' to ``$-$'' states and vice-versa, while the last term shows that the STZs are created at a rate proportional to $n_{\infty} e^{-1/ \chi}$ and annihilated at a rate proportional to their density. The plastic rate of deformation tensor is given by the rate at which STZs flip:
\beq
\label{Dpl_R}
D^{pl} = \frac{\lambda}{\tau_0} \left( R(s) n_{-} - R(-s) n_{+} \right).
\eeq
 Pechenik~\cite{Pechenik} generalized Eqs.~(\ref{number_density})~and~(\ref{Dpl_R}) to the case where the principal axes of the STZ orientation tensor $n_{ij}$ are not aligned with principal axes of the stress tensor $s_{ij}$.  These generalized equations can be written in terms of two new variables. The first is  $\Lambda \equiv n_{tot} / n_{\infty}$, where $n_{tot}$ is the tensorial generalization of ($n_{+} + n_{-}$), and corresponds to the total density of zones in a sample. The second is $ m_{ij} \equiv n_{ij} / n_{\infty}$, where $n_{ij}$ is the tensorial generalization of ($n_{+} - n_{-}$) and corresponds to the STZ orientational bias.  
  
The rate at which STZs are created/destroyed, denoted by $\Gamma$, is an important, positive quantity. Falk and Langer~\cite{Falk_L1} first proposed that $\Gamma$ should be proportional to the rate of plastic work done on the system. Pechenik~\cite{Pechenik} refined this idea, noting that the rate of plastic work could be negative since energy could be stored in the plastic degrees of freedom. He proposed an alternative definition for $\Gamma$: the rate at which energy is dissipated per STZ, which is always positive. We adopt this important assumption in our STZ model.

  In previous papers,~\cite{Pechenik,Langer_Eff}, the functional form of $\Gamma$ was necessarily complicated because thermally activated switching can release energy stored in the plastic degrees of freedom. As noted by Bouchbinder et. al~\cite{Bouchbinder}, the functional form of $\Gamma$ is considerably simplified in an athermal description because an STZ can never flip in a direction opposite to the direction of applied stress.  Therefore no energy can be stored in the plastic degrees of freedom. The rate at which energy is dissipated per STZ is the rate at which plastic work is done on the system, $Q$, divided by the volume (area in 2D) density of STZs, $\epsilon_0 \Lambda$, where $\epsilon_0 = \lambda n_{\infty}$. In terms of these new variables, the rate of deformation tensor can be written as:
\beq
\label{Dpl_C}
D^{pl}_{ij} = \frac{\epsilon_0}{\tau_0} {\cal C}(\overline{s})( \frac{s_{ij}}{\overline{s}} -m_{ij}) \Lambda,
\eeq
and the energy dissipated per STZ is
\begin{eqnarray}
\label{Gamma1}
\Gamma &=& \frac{Q}{\epsilon_0 \Lambda} \nonumber \\
 &=& \frac{ s_{ij} D^{pl}_{ij}}{\epsilon_0 \Lambda}, 
\end{eqnarray}
where $\overline{s}^{2} = 1/2 \:  s_{ij} \: s_{ij}$, and  ${\cal C}(\overline{s}) = 1/2 \left( R(\overline{s}) + R(-\overline{s}) \right)$. Based on physical considerations, ${\cal C}(s)$ is symmetric function of the stress that approaches zero as the stress approaches zero and approaches a line with a slope of unity as the stress becomes large. Bouchbinder et al.~\cite{Bouchbinder} describe a one-parameter family of functions with the correct properties, and we will use the simplest of these: ${\cal C}(s) = -2 + |s| + \exp(-|s|)(2 +|s|)$. 

Together, Eqs.~(\ref{Dpl_C}),(\ref{Gamma1}), and the tensorial generalization of Eq.~(\ref{number_density}) describe the evolution of the density of STZs and the orientational bias of STZs. The density $\Lambda$ is driven towards a Boltzmann distribution as energy is dissipated in the system:
\beq
\dot{\Lambda} = \Gamma \left( e^{- 1/ \chi} - \Lambda \right).
\eeq
The orientational bias of the STZs $m_{ij}$ is incremented to accommodate plastic strain, and reset as energy is dissipated in the system:
\bea
\label{m_eom}
\frac{{\cal D} m_{ij}}{{\cal D}t} &=& \frac{2}{\tau_0}{\cal C}(\overline{s})(\frac{s_{ij}}{\overline{s}}  - m_{ij})   - \Gamma m_{ij} \frac{e^{-1/ \chi}}{\Lambda}.
\eea
 To close the system of equations we require an equation of motion for the effective temperature, $\chi$. As mentioned in the introduction, we postulate that the effective temperature describes the energy in the disordered configurational degrees of freedom.  Ono et al.~\cite{Ono} show that at low strain rates the system is driven towards a specific steady state of disorder that corresponds to a specific value for the steady state effective temperature, $\chi_{\infty}$.

It is assumed that the effective temperature is proportional to the potential energy per particle, with a constant specific heat $c_0$. This implies that the steady state effective temperature is also proportional to the energy dissipated by the configurational degrees of freedom in steady state. Energy balance requires that the effective temperature change at a rate proportional to the rate at which energy is dissipated in the system. However, this change in configurational energy must be limited as the effective temperature approaches its steady state.  As $\chi$ approaches $\chi_{\infty}$, the heat generated by plastic work is dissipated by other mechanisms, which we do not track. Additionally, we assume that the configurational energy can be transferred between neighboring particles, so that there is a flux in the effective temperature which is proportional to its gradient. The resulting equation for effective temperature is:
\beq
\label{chi1}
c_0 \dot{\chi} = Q \left( \chi_{\infty} - \chi \right) + D_{\chi} \frac{\partial^2 \chi}{\partial y^2},
\eeq
 where $D_{\chi}$ is a diffusion constant with units of area per unit time.

 In the athermal STZ formulation there is no mechanism for relaxation  of the effective temperature to the thermal temperature $T$, so that the effective temperature everywhere tends towards $\chi_{\infty}$.  This is in contrast to a thermal formulation that permits the effective temperature to relax in regions undergoing minimal plastic deformation.  In Lennard-Jones and bulk metallic glasses well below the thermal glass temperature we expect any relaxation to be very slow, so that the athermal description is a good approximation.  However, even very slow relaxation processes may have a significant impact on localization and we re-evaluate the athermal assumption in Section~\ref{Conclusions}.

\subsection{STZ equations in an infinite strip geometry}
\label{InfiniteStrip}
We will now derive the STZ equations of motion in the geometry of a two-dimensional infinite strip driven by the boundaries in simple shear. Let $x$ be the direction of the velocity at the boundaries, and $y$ be normal to these boundaries. The speed at the boundaries is $V_0$ and the width is $L$. Therefore the average strain rate inside the material is $\overline{\dot{\gamma}} \equiv V_0 / L $, while the local strain rate is denoted $\dot{\gamma} \equiv \partial v_x / \partial y \equiv 2 D^{total}_{xy}$.  Assuming that there are no pressure gradients, Eq.~(\ref{Dtotal}) requires that the pressure field remain constant in time. 

The stress $s$ and STZ bias $m$ are traceless tensors. Let the off-diagonal elements of the tensors be denoted by $s$ = $s_{xy}= s_{yx}$ and $m$ = $m_{xy} = m_{yx}$ and the diagonal elements by $s_0$ = $s_{xx}= -s_{yy}$ and $m_0$ = $m_{xx} = -m_{yy}$.   By symmetry, all fields are constant in $x$, and  Eq.~(\ref{stat_eq}) shows the stress tensor is constant in the $y$ direction as well.  
 Substituting this into Eq.~(\ref{Dpl_C}) we find that the plastic rate of deformation tensor then has two independent components:
\bea
D^{pl} &=& \frac{ \epsilon_0}{\tau_0}{\cal C}(\overline{s})(\frac{s}{\overline{s}} - m) \Lambda ; \\
D^{pl}_0 &=& \frac{ \epsilon_0}{ \tau_0}{\cal C}(\overline{s})(\frac{s_0}{\overline{s}} - m_0) \Lambda,
\eea
and the energy dissipation per STZ, $\Gamma$, is given by:
\bea
\Gamma &=& 2  \frac{ D^{pl}_0 s_0 + D^{pl} s }{\epsilon_0 \Lambda} \nonumber \\
    &=& \frac{2}{\tau_0} {\cal C}(\overline{s})(\overline{s} - s_0 m_0 - sm).
\eea
 Because $s_{ij}$ is independent of position, the tensor equation of motion for the stress~(Eq.~(\ref{Dtotal})) can be integrated in $y$ over the interval  $ [ y = -L, \; y = +L ]$. This results in simplified equations for $s$ and $s_0$:
\bea
\dot{s} &=& \frac{\mu}{\overline{\mu}} \left( \frac{V_0}{L} - \frac{2 \epsilon_0}{\tau_0}{\cal C}(\overline{s})(\frac{s}{\overline{s}} - m)\overline{\Lambda} \right)- s_0 \frac{V_0}{L} , \\
\dot{s_0} &=& - 2 \frac{\mu}{\overline{\mu}} \left( \frac{2 \epsilon_0}{\tau_0}{\cal C}(\overline{s})(\frac{s_0}{\overline{s}} - m_0)\overline{\Lambda} \right) + s\frac{V_0}{L},
\eea
where $ \overline{ \left( \cdot \right)}$ denotes an average over a field in the $y$-direction. For example, $\overline{\Lambda}$ is:
\beq
\overline{\Lambda} = \frac{1}{2 L} \int_{-L}^{L} \Lambda (y) dy.
\eeq
and $\overline{D^{total}_{xy}} = V_0 / (2 L)$.  Eq.~(\ref{m_eom}) can be simplified to equations for $m$ and $m_0$ as follows:
\bea
\dot{m} + m_0 \dot{\gamma} &=&  - \Gamma m \frac{e^{-1/ \chi}}{\Lambda} + \frac{2}{ \tau_0} {\cal C}(\overline{s})(\frac{s}{\overline{s}} - m); \\
\dot{m_0} - m \dot{\gamma} &=&  - \Gamma m_0 \frac{e^{-1/ \chi}}{\Lambda} + \frac{2}{ \tau_0} {\cal C}(\overline{s})(\frac{s_0}{\overline{s}} - m_0).
\eea
In slowly sheared experiments and simulations, the strain rate $\dot{\gamma}$ is always much smaller than the inherent attempt frequency $1/ \tau_0$. Therefore the complicated rotation terms, $m_0 \dot{\gamma}$ and $m \dot{\gamma}$, are very small and can be neglected. We rewrite the equations of motion so that times are in units of the inverse average strain rate, $1 / \overline{\dot{\gamma}} = L/ V_0$, and lengths are in units of $L$. Note that stresses are already in units of the yield stress $\overline{\mu}$.

The resulting system of equations is given by:
\bea
\label{simple_s}
\dot{s} &=& \mu^{*} \left( 1 - \frac{2 \epsilon_0}{q_0}{\cal C}(\overline{s})\left( \frac{s}{\overline{s}} - m \right)\overline{\Lambda} \right)- s_0; \\
\label{simple_s0}
\dot{s_0} &=& - \mu^{*} \left( \frac{2 \epsilon_0}{q_0}{\cal C}(\overline{s})\left( \frac{s_0}{\overline{s}} - m_0 \right)\overline{\Lambda} \right) + s; \\
\label{simple_m}
\dot{m} &=&  \left( - \Gamma m \frac{e^{-1/ \chi}}{\Lambda} + \frac{2}{q_0}  {\cal C}(\overline{s})(\frac{s}{\overline{s}} - m)  \right); \\
\label{simple_m0}
\dot{m_0} &=& \left( - \Gamma m_0 \frac{e^{-1/ \chi}}{\Lambda} + \frac{2}{q_0}  {\cal C}(\overline{s})(\frac{s_0}{\overline{s}} - m_0)  \right); \\
\label{simple_Lambda}
\dot{\Lambda} &=& \Gamma \left( e^{- 1/ \chi} - \Lambda \right); \\
\label{simple_chi}
\dot{\chi} &=& \frac{1}{c_0} \Gamma \epsilon_0 \Lambda  \left( \chi_{\infty} - \chi \right) + D_{\chi} \frac{\partial^2 \chi}{\partial y^2},
\eea
where
\beq
\Gamma = \frac{2}{q_0} {\cal C}(\overline{s})(\overline{s} - s_0 m_0 - sm).
\eeq
 As noted in~\cite{Bouchbinder} the density of STZs, $\epsilon_0 \Lambda$, is necessarily small.  Eqs.~(\ref{simple_s}),~(\ref{simple_s0})~and~(\ref{simple_chi}) each contain this factor in their numerators and they equilibrate very slowly compared to $m$, $m_0$ and $\Lambda$, which are governed by Eqs.~(\ref{simple_m}),~(\ref{simple_m0})~and~(\ref{simple_Lambda}), respectively.  Therefore we replace $\Lambda$, $m$ and $m_0$ by their steady state values:
\bea
\Lambda &=& \Lambda_{ss}(\chi) = e^{-1/ \chi}; \\
m &=& m(s,s_0) = \left\{ \begin{matrix} s / \overline{s}, \quad \overline{s} \leq 1, \\ s / \overline{s}^{2}, \quad \overline{s} > 1, \end{matrix} \right.  \\
m_0 &=& m_0(s,s_0) = \left\{ \begin{matrix} s_0 / \overline{s}, \quad \overline{s} \leq 1, \\  s_0 / \overline{s}^{2}, \quad \overline{s} > 1. \end{matrix} \right. 
\eea
 Below the yield stress the solid deforms only elastically because all the existing STZs are already flipped in the direction of stress. Three simple partial differential equations remain:
 
\begin{eqnarray}
\label{simple_s2}
\dot{s} &=& \mu^{*} \left( 1 - \frac{2 \epsilon_0}{q_0}{\cal C}(s)\left(\frac{s}{\overline{s}} - m \right)\overline{\Lambda} \right)- s_0, \\
\label{simple_s02}
\dot{s_0} &=& - \mu^{*} \frac{2 \epsilon_0}{q_0}{\cal C}(s)\left(\frac{s_0}{\overline{s}} - m_0 \right)\overline{\Lambda} + s, \\
\label{simple_chi2}
\dot{\chi} &=& \frac{2 \epsilon_0 {\cal C}(s)}{c_0 q_0} \left( \overline{s} - s_0 m_0 - s m \right)  e^{-1/ \chi} ( \chi_{\infty} - \chi) \nonumber \\
 & & + \: D^{*}_{\chi} \frac{\partial^2 \chi}{\partial y^2}.
\end{eqnarray}

This set of equations has several dimensionless parameters. The volume (area in 2D) fraction of the strip covered by STZs or the probability of finding an STZ at a given point in the strip is $\epsilon_0 \Lambda$.  $\Lambda$ is the probability that a particle participates in an STZ. Assuming that STZs have a characteristic size of a few particle diameters,  $\epsilon_0 = \lambda n_{\infty}$ is the characteristic number of particles in an STZ.  The parameter $\mu^{*} = \mu/ \overline{\mu}$ is the ratio of average material stiffness to STZ stiffness, and corresponds to the slope of the linear elastic portion of a stress-strain curve.  The specific heat in units of the STZ formation energy, $c_0$, determines how fast the effective temperature changes compared to plastic rearrangements.  The parameter $q_0 = \tau_0 V_0  / L$ is the ratio of the two physical timescales in the system, the inverse STZ attempt frequency and the average inverse strain rate. Finally, $\chi_{\infty}$ is the steady state effective temperature in units of the STZ formation energy divided by the Boltzmann constant. 

The effective temperature diffusivity is the remaining parameter and requires some discussion. The dimensionless parameter $D^{*}_{\chi}$ in Eq.~(\ref{chi1}) can be written as follows:
\beq
D^{*}_{\chi} = (D_{\chi}/ c_0)(L / V_0) (1/L^2),
\eeq
where $D_{\chi} = a^{2} / \tau_{D}$ is a diffusivity with units of length scale squared per unit time. In STZ theory there are two candidates for the attempt frequency $1 / \tau_{D}$: the local strain rate $\dot{\gamma}_{loc}$ and the inherent material attempt frequency $1 / \tau_{0}$.  The first possibility implies that the local potential energy diffuses at a rate proportional to the rate of local particle rearrangements, while the second implies that the energy flux depends on an inherent vibrational/frictional timescale.

 Naively, one might think that the inherent frequency $1/ \tau_{0}$ is too high and results in a diffusion constant much too large to be physically reasonable.  However, if one assumes that configurational energy diffusion is an activated process with large energy barriers ($\tau_{D} = \tau_0 \exp( \Delta E / k T)$), then the resulting diffusion time $\tau_D$ could be large and yet remain independent of strain rate.

 We have examined solutions to the STZ equations with an effective temperature diffusivity $D^{*}_{\chi}$ (a)  proportional to the local strain rate and (b) independent of strain rate. In case (a) multiple shear bands are more likely, the effective temperature takes much longer to reach its peak value, and the dynamics are more complicated.  In case (b) a single shear band is more likely and the effective temperature quickly reaches its plateau. While further study is required to determine which diffusion frequency best describes the data, an effective temperature diffusivity that is independent of strain rate results in simpler dynamic equations, and we will use this assumption for the remainder of this paper. 

\section{Numerical integration of the STZ equations}
\label{Numerical} 
  In this section we show that numerical integration of the simple, macroscopic STZ equations (Eqs.~(\ref{simple_s2}-\ref{simple_chi2})) qualitatively and quantitatively match molecular dynamics simulations of a Lennard-Jones glass by Shi, Katz, Li and Falk~\cite{Shi}.  The simulations were performed in a simple shear geometry with periodic Lees-Edwards boundary conditions. The parameter range for the simulations was chosen to ensure that the glass exhibits strain localization, and stress-strain curves were computed for a variety of quenches and strain rates. 
 
The investigators calculated the potential energy per atom and strain rate as a function of position for several different strains. They used the data to investigate the hypothesis (found in the STZ model and other models) that the local plastic strain rate is related to an effective temperature or free volume. In~\cite{Shi}, they show that a value for the steady state effective temperature,$\chi_{\infty}$, can be rigorously extracted from a thorough analysis of the simulation data. In this paper we numerically integrate the STZ equations in the same infinite strip geometry, and compare the results to simulation data.  Building on the work of Shi, {\em et al.}, we choose parameters for the STZ theory to match the conditions found in the simulations.

In experiments and simulations, boundary conditions on the particles at the top and bottom of the strip impact the dynamics and may influence strain localization.  Shi, {\em et al.} chose periodic boundary conditions for their simulations, while other investigators~\cite{Varnik} have chosen rigid rough walls. While both sets of simulations exhibit strain localization, features of the localization process are different in each case and it is unclear whether the differences are due to the dissimilar boundary conditions or other features of the simulations.

 Similarly, the STZ equations require boundary conditions for the effective temperature field at the top and bottom of the strip.  We have studied the STZ equations with both periodic and no conduction ($\partial \chi / \partial y = 0$) boundary conditions and found qualitatively and quantitatively similar results. For the remainder of this paper we will describe solutions to the STZ equations with periodic boundary conditions for comparison to the results of Shi, {\em et al.}

\subsection{Matching parameter values to simulations}

 The first task is to choose values for the parameters in STZ theory consistent with those in the simulations by Shi, {\em et al.} The simulation data for the stress-strain curves at a particular strain rate can be fit by a range of values for $\chi_{\infty}$, $q_0$, and the specific heat $c_0$. However, a much narrower range of parameters fit both the stress-strain data and the data for the strain-rate as a function of position.

We estimate an order of magnitude for each parameter motivated by physical considerations. Table~\ref{par_table} lists the specific parameter values used in the numerical integration.   The steady state effective temperature $\chi_{\infty}$ determines the steady state density of STZs: $\Lambda_{\infty} = \exp[-1 / \chi_{\infty}]$. To ensure that STZs are rare, $\chi_{\infty}$ should be around $0.1$. Shi, {\em et al.}, extracted $\chi_{\infty} = 0.15$ from the data using the hypothesis that the strain rate is proportinal to $e^{-1/\chi}$, and we will use this value here. The parameter $\epsilon_0$ is the characteristic number of particles in a shear transformation zone. Studies of non-affine particle motion in a 2D Lennard-Jones glass~\cite{Falk_L1} and a 2D system of hard spheres~\cite{Gregg} suggest that an STZ has a radius of a few particle diameters. This implies that the number of particles in an STZ is of order $10$ for 2D simulations.  Because $c_0$ is a dimensionless constant and does not depend on other scales in the problem, $c_0$ should be of order unity. The parameter $\mu^{*}$ is the slope of the linear elastic part of the stress-strain curve plotted in units of the yield stress. It corresponds to the ratio of the elastic material stiffness to the STZ stiffness, and should be much greater than one. In the MD simulations, $\mu^{*}$ is about 70.  
\begin{table}
\begin{tabular}{|c|c|}
\hline
{\bf Parameter} & {\bf Value} \\ \hline
$\chi_{\infty}$ & 0.15 \\ \hline
$\epsilon_0$ & 10 \\ \hline
$c_0$ & 1 \\ \hline
$D^{*}_{\chi}$ & 0.01 \\ \hline
$\mu^{*}$ & 70 \\ \hline
$q_0$ & $1 \times 10^{-6}$ \\ \hline
$\chi_0$ & $0.68,0.69,0.74$ \\ \hline
\end{tabular}
\caption{ \label{par_table} List of parameter values used in the numerical integration of the STZ model, Eqs.~(\ref{simple_s3},\ref{simple_chi3}).}
\end{table}

 Eqs.~(\ref{simple_s2}-\ref{simple_chi2}) can be further simplified because $\mu^{*}$ is large. For $\overline{s} \leq 1$, the rate of plastic deformation is zero and  and Eq.~(\ref{simple_s2}) indicates that the stress increases proportionally to $\mu^{*}$. During this same time, Eq.~(\ref{simple_s02}) requires that $s_0$ increases proportionally to $s$, which is less than unity.  During the linear elastic response, the off-diagonal component of the stress is order $\mu^{*}$ larger than the on-diagonal stress, which matches our physical intuition for simple shear. Numerical integration of Eqs.~(\ref{simple_s2}-\ref{simple_chi2}) confirms that $s_0$ remains about two orders of magnitude smaller than $s$. We therefore use the approximation $\overline{s} = s$ and $s_0 = 0$. This results in two simple equations for the STZ dynamics:
\bea 
\label{simple_s3}
\dot{s} &=& \mu^{*} \left( 1 - \frac{2 \epsilon_0}{q_0}{\cal C}(s)\left( 1 - m(s)\right)\overline{\Lambda} \right), \\
\label{simple_chi3}
\dot{\chi} &=& \frac{2 \epsilon_0 {\cal C}(s) \: s}{c_0 q_0} \left( 1  - m(s) \right)  e^{-1/ \chi} ( \chi_{\infty} - \chi) \nonumber \\
 & & + \: D^{*}_{\chi} \frac{\partial^2 \chi}{\partial y^2}.
\eea

 The remaining parameters are  $D^{*}_{\chi}$ and $q_0$. As mentioned in the previous section, we postulate that the effective temperature diffusion $D^{*}_{\chi}$ is a constant independent of strain rate which we determine from the long-time diffusion of the shear band in the simulations.  We find that $D^{*}_{\chi} \simeq 0.01$ matches the simulation data.

  The Lennard-Jones glass has a natural time scale $t_0 = \sigma_{SL} \sqrt{M/ \epsilon_{SL}}$, where $\sigma_{SL}$ is the equilibrium distance between small and large particles, $\epsilon_{SL}$ is the depth of the potential energy well between the two species and $M$ is the mass. The average strain rate in the simulations varies from $2 \times 10^{-5} t_0^{-1}$ to $5 \times 10^{-4} t_0^{-1}$. 

If we assume that the STZ attempt frequency $\tau_0$ is approximately $t_0$, then $q_0 \simeq 2 \times 10^{-5}$ and the resulting numerical solutions to the STZ equations never exhibit localization. This is inconsistent with simulation data. In order to match the localization seen in the simulations, we are required to choose a value for $\tau_0$ that is an order of magnitude smaller than $t_0$. We find that $q_0 \simeq 1 \times 10^{-6}$ results in STZ solutions where strain rate matches results from the MD simulations. We will discuss the implications of this in Section~\ref{long_time}.

  Initial conditions for $s$ and $\chi$ must also be determined. Because the stress is constant as a function of position and the simulations begin in an unstressed state, $s(t=0) = 0$.  The initial effective temperature $\chi(y, t=0)$ is a {\em function} of position with many more degrees of freedom. In the MD simulations, the initial potential energy is nearly constant as a function of position with fluctuations about the mean. Therefore, the initial effective temperature is also nearly constant with small amplitude perturbations about its mean. We expect fluctuations in the effective temperature to occur on the scale of a few particle radii in dimensionless units. The simulation box of Shi, {\em et al.} has a width of about 300 particle radii, so perturbations which span five particle radii have a nondimensionalized wavelength $w = 1/60$.

 The amplitude of the initial perturbations, $\delta \chi_0$, can be approximated empirically from the initial potential energy per atom as a function of position, shown in Figure~\ref{position_data}(b).  The potential energy at 0\% strain exhibits small perturbations with standard deviation $\sim 0.02 \: \epsilon_{SL}$ about a mean value of $-2.51 \: \epsilon_{SL}$, while the system reaches a maximum of $-2.42 \: \epsilon_{SL}$ in the shear band at larger strains. Assuming the maximum potential energy in the band corresponds to $\chi_{\infty}$ and the initial mean potential energy per atom corresponds to $\chi_0$, the amplitude of initial perturbations is about $\delta \chi = 0.02$. This is consistent with a thermodynamic calculation~\cite{Landau} for the magnitude of fluctuations about the effective temperature, $< (\delta \chi)^{2}> = \chi^{2} / (c_0) \simeq 0.01$. 

 We use two different methods for generating an initial effective temperature distribution for the STZ equations. The first is to use a deterministic function with a single peak that serves as a nucleation point for shear bands:  $\chi(y, t=0) = \chi_0 + \delta \chi_0 \: \mbox{sech} (y / w)$. The second method generates a random number from a uniform distribution, and smooths those values using a simple moving average of width $w$. The resulting function is normalized so that its standard deviation from the mean value is $\delta \chi_0$, and the mean is set at $\chi_0$. 

The first type of initial condition is less physically realistic but more tractable because it generates at most a single shear band.  The second type of initial condition can generate solutions with varying numbers of shear bands, depending on the system parameters.  We find that for the parameter range which best fits the simulation data, the STZ equations with random initial conditions generate a solution with a single shear band, which agrees with observations from the simulations and further validates our choice of STZ parameters.

\subsection{Comparison of macroscopic stress-strain behavior}
 
\begin{figure}[h]
\centering \includegraphics[height=5.5cm]{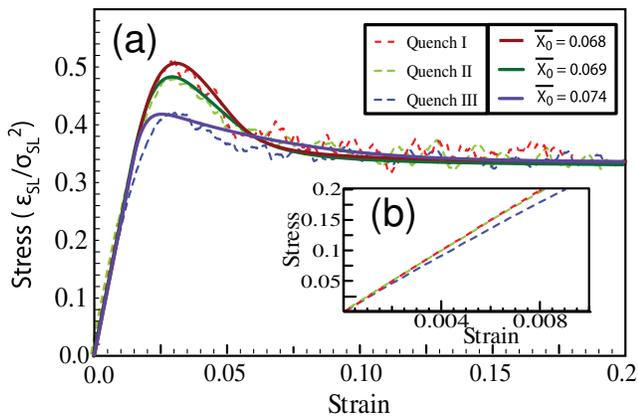}
\caption{\label{match_stress_falk} (color online) (a)Simulation data and STZ theory for stress vs. strain from Ref.~\cite{Shi}. Dashed lines are simulation data for three different initial preparations. The red curve corresponds to a sample quenched for the longest time (100,000 $t_0$)before the sample was sheared and has an initial average potential energy per atom of -2.507. The green and purple curves were quenched for 50,000 $t_0$ and 10,000 $t_0$ respectively, and have initial PE/atom of -2.497 and -2.477. The solid lines correspond to STZ solutions with values for the initial effective temperature ($\chi_0$) that best fit the simulation data: 0.062, 0.063 and 0.067, respectively. (b)Linear elastic regime of the simulation data for the three different quenches.}
\end{figure}

  We numerically integrate Eqs.~(\ref{simple_s3},~\ref{simple_chi3}) with parameters values discussed above, for many different values of $\chi_0$. Figure~\ref{match_stress_falk}(a) is a plot of stress vs. strain for three different initial preparations of a material starting from rest and driven in simple shear. Dashed lines correspond to MD simulation data from samples permitted to relax for three different amounts of time before being sheared.   Solid lines are numerical solutions to the simplified STZ model (Eqs.~(\ref{simple_s3})~and~(\ref{simple_chi3})) for three different average initial effective temperatures. 

All of the samples (for both the MD simulations and the STZ solutions) exhibit strain localization, which influences the stress-strain curves shown in Figure~\ref{match_stress_falk}. We will discuss how localization affects the macroscopic stress-strain curves later in this section, and analyze the localization process in Section~\ref{Stability}.
 
The STZ solutions shown differ only in the initial mean value for the effective temperature, $\chi_0$, and are the best least-squares fit as a function of $\chi_0$ for each quench. Note that general features of the stress-strain curves match: there is a linear elastic segment, followed by a decrease in the stress slope as the material begins to deform plastically, a peak at about 2 \% strain, and stress relaxation as the material softens and stored elastic energy is released.  

The glass in the MD simulations behaves differently depending on how long the system was quenched before shearing, and the STZ model captures this behavior.  A longer quench results in a more ordered solid and corresponds to a lower initial effective temperature. Numerical integration of the STZ equations indicate that lower initial effective temperatures generically result in higher peak stresses. This matches both the MD simulations and physical intuition:  a more ordered solid takes more time and stress to plastically deform because more STZs must be created to permit the deformation. 

 In Figure~\ref{match_stress_falk}(b) the slope ($\mu^{*}$) of the linear elastic segment of the most disordered glass (Quench III) is smaller than in the samples that were quenched for longer times before shearing. A linear fit to these data shows that $\mu^{*} \simeq 70$ for Quenches I and II, while $\mu^{*} \simeq 60$ for Quench III.

 We model the material surrounding the STZs as an elastic medium, and the parameter $\mu^{*}$ is the ratio of the elastic material stiffness to the STZ stiffness.  We assume that the STZ stiffness is constant for different sample preparations.  Therefore a variation in $\mu^{*}$ between samples indicates that the surrounding elastic medium is less stiff in the more disordered material. This is consistent with the work of Maloney and Lema{\^\i}tre, who have shown that the elastic shear modulus is smaller in more disordered materials due to non-affine particle motion~\cite{Maloney}.

 Although we could better fit the MD simulation data by allowing $\mu^{*}$ to vary between samples, in order to limit the adjustable parameters in the theory we fix $\mu^{*} = 70$ for all samples. As expected, Figure~\ref{match_stress_falk}(a) shows that the numerical STZ results closely match the simulation data for Quenches I and II, while the best STZ fit systematically deviates from the simulation data for Quench III.

\subsection{Comparison of strain localization inside the material}

   One feature of the infinite strip geometry is that it permits the system to achieve very large strains without fracturing, so that both theory and simulation can track system evolution over very large strains. The STZ solutions not only match the short-time macroscopic stress-strain behavior, but also match the long-time dynamics of strain localization within the strip.  

\begin{figure}[h]
\centering \includegraphics[height=11.0cm]{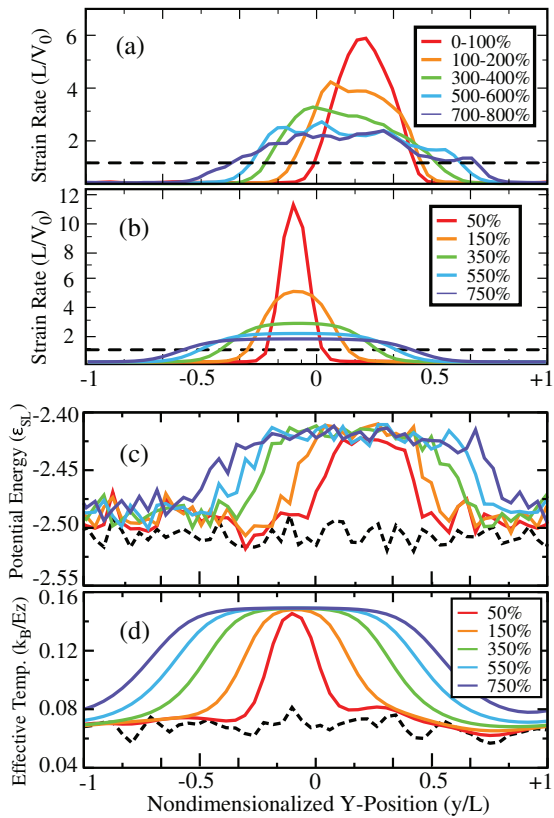}
\caption{\label{position_data} (color online) {\bf (a)} Simulation data for strain rate vs. position at various strains from Ref.~\cite{Shi}. Strain rates are averaged over bins of 100 \% strain. The dashed line corresponds to the imposed average strain rate. {\bf (b)} Theoretical STZ solutions for strain rate vs. position at various strains. Strain rate is evaluated for specific values of the strain corresponding to the midpoint of each binning range for simulation data. The dashed line is the average strain rate. {\bf (c)} Simulation data for potential energy as a function of position~\cite{Shi} and {\bf (d)} Theoretical solution for the effective temperature as a function of position. The dashed lines correspond to the initial values of the potential energy per atom and effective temperature.}
\end{figure}

Figure~\ref{position_data} shows the strain rate as a function of position for various values of the strain (corresponding to different times) for (a) the MD simulations and (b) STZ theory. This figure also shows (c) the simulation potential energy per atom  and (d) theoretical effective temperature as functions of position. The simulation data is averaged over increments of 100 \% strain, while the STZ model is evaluated for specific values of the strain corresponding to the midpoint of each binning range. 

Localization is evident in both the simulation data and numerical STZ solutions. The strain rate as a function of position inside the material exhibits very slow relaxation over 800 \% strain. This is in contrast to the much faster stress dynamics that attain steady state in less than 10\% strain, as seen in Fig.~\ref{match_stress_falk}. In the numerical STZ solutions, the effective temperature attains a maximum in the same physical location within the strip as the strain rate.  This is remarkably similar to the dynamics of the potential energy per atom in the simulations, and completely consistent with the assumption that $\chi$ is proportional to the potential energy. Shi, {\em et al.} were the first to systematically check this hypothesis and they used it to extract various STZ parameters, such as $\chi_{\infty}$~\cite{Shi}.

\subsection{Implications for constitutive laws}
\begin{figure}[h]
\centering \includegraphics[height=8cm]{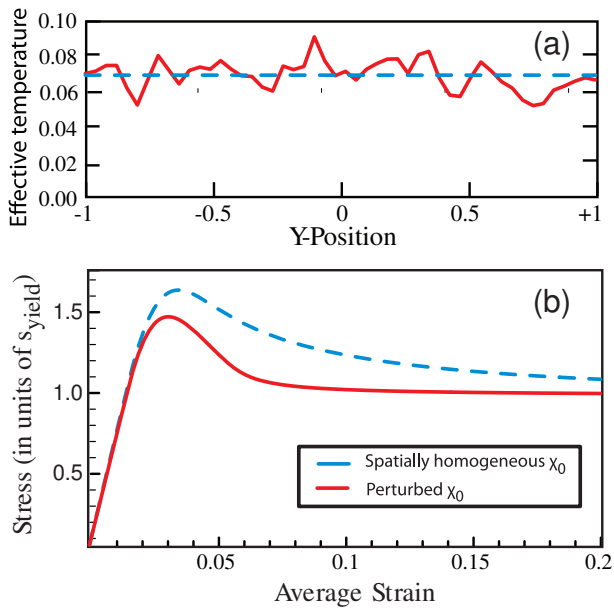}
\caption{\label{compare_stress} (color online) Macroscopic stress-strain curves are sensitive to localization. {\bf (a)} The initial effective temperature distribution is either homogeneous (blue) or slightly perturbed (red), though the average value $\chi_0$ is the same in both cases. {\bf(b)} The stress vs. strain predictions are significantly different for these two different initial conditions. The perturbed system localizes, as shown in Figure~\ref{position_data}(d), and this changes the resulting macroscopic behavior. }
\end{figure}

   Strain localization has a large effect on the short-time macroscopic stress-strain behavior of the system.  For example, Fig.~\ref{compare_stress} (a) shows two possible initial conditions for the initial effective temperature field. The average initial effective temperature is the same in both cases.  One initial condition for the effective temperature varies with position while the other is constant as a function of position. Figure~\ref{compare_stress} (b) shows the resulting stress-strain curves for each case. When $\chi_0$ varies as a function of position, the system localizes and the resulting stress-strain curve does not reach as high a peak value, because the material releases elastic energy more quickly.  The steady state stress is the same in each case. This is just one example of the general sensitivity of macroscopic state variables to microscopic details.

  The fact that localization influences the macroscopic stress-strain curves has important implications for constitutive laws. Constitutive laws provide a relationship between the strain rate, stress, and a set of state variables that characterize the internal structure of a material interface. They describe how features of microscopic particle dynamics determine macroscopic frictional properties, and are used extensively in models of earthquakes~\cite{Dieterich,Ruina}, granular flows~\cite{Jop}, and machine control~\cite{urbakh}. Several investigators have adapted STZ theory to describe the macroscopic behavior of lubricated surfaces~\cite{Anael2}, dense granular flows~\cite{Gregg}, and earthquake faults~\cite{Daub}, but all of these formulations assume that the STZ variables do not vary as a function of position across the sheared interface. Deriving new constitutive laws based on localized dynamics is beyond the scope of this paper. However, our analysis shows that localization is possible in STZ models and that the corresponding constitutive behavior is altered. Therefore it is important for localization to be included in modeling efforts, and this will be a direction of future research.

\section{Stability analysis} 
\label{Stability}
Continuum models for amorphous solids create a framework for understanding the instability that leads to localization in amorphous materials. In the previous section we showed that STZ theory with effective temperature exhibits strain localization. In the following subsections we investigate the stability properties of the dynamical system given by Eqs.~(\ref{simple_s3})~and~(\ref{simple_chi3}).

\subsection{Steady state linear stability}
\label{steady_state}
  The first step in understanding how the simplified STZ model given by Eqs.~(\ref{simple_s3})~and~(\ref{simple_chi3}) responds to perturbations is to perform a linear stability analysis around the steady state solution. Setting $\dot{s} = 0$ in Eq.~(\ref{simple_s3}), results in the following equation for the steady state flow stress $s_f$:
\beq
\label{sf}
1 = \frac{2 \epsilon_0}{q_0}{\cal C}(s_f)(1 - 1/s_f)\exp[ -1/ \chi_{\infty}].
\eeq
 For the values of $q_0$ and $\chi_{\infty}$ in Table~\ref{par_table}, $s_f = 1.005$ --- the yield stress is nearly equal to the steady state stress. The steady state solution for $\chi$ is spatially homogeneous:  $\chi(y) = \chi_{\infty}$.

We now show that the perturbed system, $s = s_f + \widetilde{s}, \chi = \chi_{\infty} + \delta \chi(y)$, is linearly stable.
To simplify notation, we define the functions $f$ and $g$:
\begin{eqnarray}
f(s ,\chi) &\equiv& \frac{\partial s}{\partial t}  \nonumber \\
 &=& \label{def_f}
\mu^{*} \! \! \left( 1 - 2 \frac{\epsilon_0 {\cal C}(s)}{q_0} (1 - m)\! \int_{-L}^{L} \frac{ e^{-1 / \chi}}{2L} dy \right) \! ; \\
g(s , \chi) &=&  \frac{\partial \chi}{\partial t} \nonumber \\
 &=& \label{def_g} \frac{2 s \epsilon_0 {\cal C}(s)}{q_0 c_0} (1 - m)e^{-1 / \chi} \left( \chi_{\infty} - \chi \right) \nonumber \\
 & &  + \: \: {\cal D}_{\chi} \frac{\partial^2 \chi}{\partial y^2}.
\end{eqnarray}
 The perturbation to $\chi$ can be written as a sum of normal modes that satisfy the boundary conditions:
\beq
\label{normal_modes}
\delta \chi(y) = \sum_{k=- \infty}^{\infty} \delta \chi_{k} e^{ i k y}, \quad \quad k = \frac{n \pi}{L}. 
\eeq
The operator $\partial^2 / \partial y^2$ is diagonalized in the basis of normal modes and therefore the dynamics of each normal mode are decoupled. In the limit of infinitesimal perturbations, terms of order $\delta \chi^2$ and $\widetilde{s}^2$ can be neglected. This results in the following linear equations for each mode:
\begin{eqnarray}
\label{ss_stab1}
\dot{\widetilde{s}} &=& \frac{\partial f}{\partial s}(s_f, \chi_{\infty})  \widetilde{s} +  \frac{\partial f}{\partial s}(s_f, \chi_{\infty})\delta \chi_k,\\
\dot{\delta \chi_k} &=&  \frac{\partial g}{\partial s}(s_f, \chi_{\infty}) \widetilde{s} + \frac{\partial g}{\partial \chi}(s_f, \chi_{\infty}) \delta \chi_k.
\end{eqnarray}
The second term in Eq.~(\ref{ss_stab1}) requires some additional explanation, because the operator $\partial f / \partial s$ includes an integral. The action of this operator on a normal mode $\delta \chi_k e^{i k y}$ is given by:
\bea
\frac{\partial f}{\partial \chi} \delta \chi_k e^{i k y} &=&
{\cal F}(s_f, \chi_{\infty}) \: \frac{1}{2L} \int_{-L}^{+L} \delta \chi_k e^{ i n \pi y / L} dy \\
 &=& \left\{ \begin{array}{ll}
 0, & \quad k \ne 0, \\ 
 {\cal F}(s_f, \chi_{\infty}),  &\quad k = 0, \end{array} \right.
\eea
where 
\beq
{\cal F}(s_f, \chi_{\infty}) = -\frac{2 e^{-1/\chi_{\infty} } \epsilon_0 \mu^{*}
   \left(1-\frac{1}{s_f}\right)  {\cal C}(s_f)}{ q_0  \chi_{\infty} ^2} \delta \chi_k .
\eeq

This analysis reveals a particularly interesting and important feature of the STZ model dynamics which we will return to in the next section. There is a fundamental difference between the dynamics of spatially homogeneous perturbations to $\chi$ and perturbations with zero mean. Because the stress is always spatially homogeneous, to linear order the stress dynamics depend {\em only on the average value of $\chi$} and are completely unaffected by zero mean perturbations to $\chi$.
 
 In order to determine the linear stability of each mode, we calculate the two eigenvalues (as a function of $k$) for the steady state linear operator $A^{ss}(k)$, which is the two-by-two matrix given by:
\bea 
\begin{bmatrix} \dot{\widetilde{s}} \\ \dot{\delta \chi_k} \end{bmatrix} &\equiv& A^{ss}(k)\begin{bmatrix} \widetilde{s} \\ \delta \chi_k \end{bmatrix} \\
 &\equiv& \begin{bmatrix}  A_{11}(k)  & A_{12}(k) \\ A_{21}(k)  & A_{22}(k) \end{bmatrix}
  \begin{bmatrix} \widetilde{s} \\ \delta \chi_k \end{bmatrix}.
\eea
The two diagonal terms can be written:
\bea
A_{11} &=& \frac{2 e^{-1/\chi_{\infty} } \epsilon_{0} \mu^{*}}{q_0 } \left( \left( \frac{1}{s_f}- 1 \right){\cal C }'(s_f)- \frac{{\cal C}(s_f)}{s^2} \right)\nonumber \\
            & & \times \: \Theta(s_f-1), \\
A_{22} &=& \frac{2 \epsilon_0 s_f e^{-1/ \chi_{\infty}}} {c_0 q_0 } \nonumber \\
       & & \times \left( - {\cal C}(s_f) (1-\frac{1}{s_f })\right) \Theta (s_f-1)  -  D^*_{\chi} k^2,
\eea
where $\Theta$ is the unit step function.

If either of the two eigenvalues of $A^{ss}$ has a positive real part for a particular value of $k$, then that mode grows exponentially and the system is unstable with respect to perturbations with that wavenumber $k$.

 For $k \neq 0$, $ A_{12}(k)$ is zero and the matrix is lower triangular, so the eigenvalues are simply $A_{11}$ and $A_{22}$. Because ${\cal C}(s)$ is monotonically increasing and positive for $s > 0$, both eigenvalues are negative for all values of $k \neq 0$. The eigenvalues calculated for $k=0$ are negative also. Therefore, the STZ model is stable with respect to perturbations in steady state. An analysis by Foglia of the operator $A_{22}$ for an STZ model that included thermal effects resulted in the same conclusion~\cite{Foglia}.

\subsection{Time-varying stability analysis}
\label{time_varying}

In simulations and numerical integration of the STZ equations, localization of strain first occurs before the stress reaches a steady state. Transient localization is also seen in numerical simulations of Spaepen's free volume model by Huang et al.~\cite{Huang}, and in the Johnson-Segalman  model for complex fluids~\cite{Fielding}. This motivates us to study the stability of transient STZ dynamics. 

 The field $\chi$ in the STZ model given by Eqs.~(\ref{simple_s3},~\ref{simple_chi3}) can be rewritten as the sum of normal modes.  As discussed in Section~\ref{steady_state}, for small perturbations with wavenumber $k$, the governing equations for the $k=0$ mode are fundamentally different from all other modes. This motivates us to write the field $\chi$ at each point in time as the sum of a spatially homogeneous field $\overline{\chi}(t)$ and zero-mean perturbations:
\beq
\label{nm2}
\chi(y,t) = \overline{\chi}(t) + \sum_{k \neq 0} \delta \chi_{k}(t) e^{i k y}, \quad k = \frac{n \pi}{L}.
\eeq
To analyze the transient stability of this system, we permit the $k=0$ mode, $\overline{\chi}(t)$, to be arbitrarily large but constrain the zero mean perturbations, $\delta \chi_{k}$, to be small. This is slightly different from the normal mode analysis in Section~\ref{steady_state} where both $k=0$ and $k \ne 0$ modes were small.  Substituting Eq.~(\ref{nm2}) into Eq.~(\ref{def_f}) and neglecting the second order terms in $\delta \chi_{k}$ results in the following equation:
\bea
\label{s_dec}
\dot{s} &=& f(s ,\overline{\chi} + \sum_{k \neq 0} \delta \chi_{k}(t) e^{i k y}) \nonumber \\
 &=& \mu^{*} \left( 1 - 2 \frac{\epsilon_0}{q_0} {\cal C}(s) (1 - m_0) e^{-1 / \overline{\chi}} \right. \nonumber \\
 & & \left. \times \left( 1 + \frac{1}{\overline{\chi}^2} \sum_{k \neq 0} \frac{1}{2L}\int_{-L}^{L}  \delta \chi_{k}(t) e^{i k y}  dy \right) \right) \nonumber \\ 
 &=& \mu^{*} \left( 1 - 2 \frac{\epsilon_0}{q_0} {\cal C}(s) (1 - m_0) e^{-1 / \overline{\chi}} \right) \nonumber \\
 &=& f(s,\overline{\chi}).
\eea
This indicates that to linear order, zero-mean perturbations to $\chi$ do not affect the stress. Substituting  Eq.~(\ref{nm2}) into Eq.~(\ref{def_g}) results in the following linearized equation for each perturbation mode:
\beq
\label{chipert_dot}
\dot{\delta \chi}_k(t) =  \frac{\partial g}{\partial \chi}(s(t), \overline{\chi}(t), k) \: \: \delta \chi_k.
\eeq
This is a linearized equation for the dynamics of small perturbations about a spatially homogeneous, time varying solution. It is valid whenever the perturbations are small, even if the magnitude of $\overline{\chi}(t)$ is large.

 Physically, this corresponds to the following experiment: the system is started from a spatially homogeneous initial condition and the system remains spatially homogeneous until a small, zero-mean perturbation in introduced at time $\tau$. The time-varying linear operator $\frac{\partial g}{\partial \chi}(s(\tau), \overline{\chi}(\tau))$ describes the growth or decay of these small perturbations. The functions $s(t), \overline{\chi}(t)$ are the solutions to the STZ equations with spatially homogeneous initial conditions $s(t=0)= s_0$, $\chi(y,t=0) = \chi_0$. If the initial conditions are homogeneous, the STZ equations are simply ordinary differential equations, which are much easier to solve than the inhomogeneous equations.

This description for the linearized start-up dynamics is relevant to many experiments and simulations. For example, the simulations by Shi, {\em et al.} begin from a state where the  initial potential energy, which corresponds to the initial effective temperature, is roughly constant as a function of position. This corresponds to a spatially homogeneous initial condition for the effective temperature, $\chi (t =0, y) \equiv \chi_0 = c$.  Additionally, the MD samples are started from an unstressed state, which corresponds to $s_0 = 0$. The potential energy does vary slightly as a function of position, which corresponds to small, zero-mean perturbations to $\chi$ introduced  at time $t=0$.

We define the stability exponent $\omega_{c}(k,t)$ in the following manner. Let $s_c(t), \overline{\chi}_{c}(t)$ be the unique spatially homogeneous solution to the STZ equations starting from the initial condition $s_0 = 0, \chi_0 = c$. Then the exponent at time $\tau$ is defined as:
\bea
\label{st_exp}
\omega_{c}(k, \tau) &=& \frac{\partial g}{\partial \chi}(s_c(\tau), \overline{\chi}_c(\tau), k) \nonumber \\
 &=& \frac{2 e^{-1/\overline{\chi}_c } \epsilon_0 s_c}{c_0 q_0}\nonumber \\
 & & \times \left( \left( \frac{\chi_{\infty}-\overline{\chi}_c}{\overline{\chi}^2_c}- 1 \right) {\cal C}(s_c) (1-\frac{1}{s_c} ) \right) \nonumber \\
 & & \times \: \Theta(s_c -1)- D^*_{\chi} k^2,
\eea
where the functions $s_c(t)$ and $\overline{\chi}_c(t)$ are understood to be evaluated at $t=\tau$ and $\Theta(s)$ is the unit step function.

 This exponent describes the rate of growth or decay of a small perturbation with wavenumber $k$ introduced at time $\tau$ to the solution $s_c(t), \overline{\chi}_{c}(t)$. If the real part of $\omega_{c}(k, t)$ is greater than zero then the perturbations grow exponentially and the system is unstable, and otherwise the system is stable with respect to  perturbations.

Note that $\omega_{c}(k,t)$ contains a single term that depends on the wavenumber $k$ --- this term is proportional to $-k^2$ and corresponds to the diffusion of effective temperature. Because diffusion can only act to stabilize perturbations to $\chi$, the most dangerous mode corresponds to $k = \pm \pi / L$.   

\begin{figure}[h]
\centering \includegraphics[height=4.5cm]{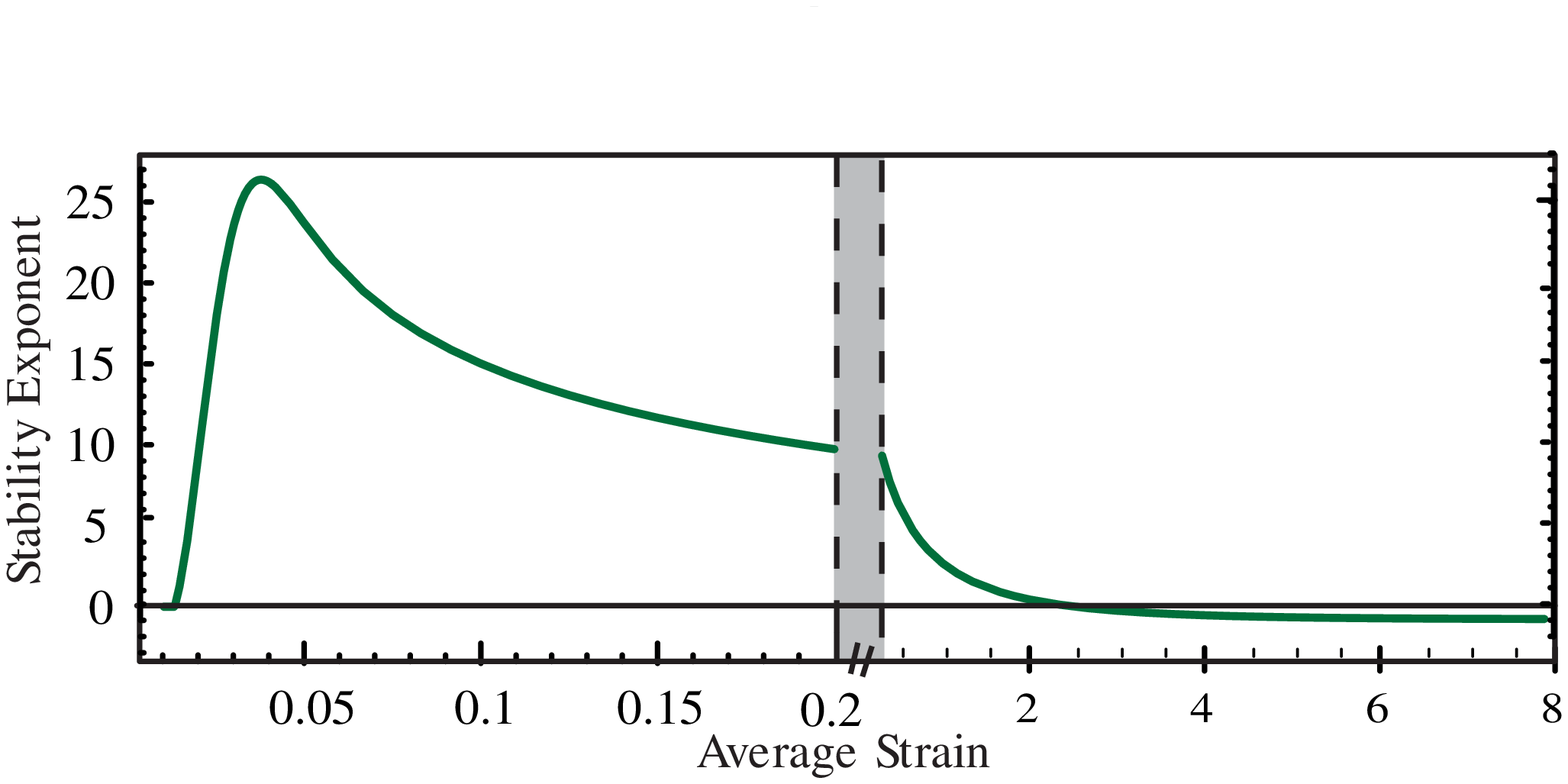}
\caption{\label{a22} The stability exponent $\omega_c(k,t)$ with $c=0.06$ and $k= \pi/L$ as a function of total strain for spatial perturbations to the effective temperature in the limit $k \rightarrow 0$. The system is linearly unstable with respect to spatial perturbations to $\chi$ when $\omega_c(k,t)$ is greater than zero, and stable otherwise.  The figure contains two scales for the average strain in order to simultaneously illustrate both the short- and long-time behavior.}
\end{figure}

 A plot of $\omega_{c}(k,t)$ for $c = \chi_0 = 0.6$ and $k = \pi / L$ is shown in Fig.~\ref{a22}. The system is marginally stable with respect to perturbations during the linear elastic response of the system, highly unstable at intermediate times, and becomes stable with respect to perturbations as $\overline{\chi}(t)$ approaches $\chi_{\infty}$.  A lower bound for stable $\chi$ is determined from  Eq.~(\ref{st_exp}) as follows. Because the diffusion term is always negative, an upper bound on the stability exponent for all values of $k$ is given by:
\beq
\omega_{c}(k, t) < H(s) e^{-1/ \overline{\chi}_c(t)} \left( \frac{\chi_{\infty}-\overline{\chi}_c(t)}{\overline{\chi}^2_c(t)-1} \right)
\eeq
where $H(s)$ is zero for $s \leq 1$ and positive for $s > 1$. Therefore the stability exponent is less than zero when $\overline{\chi}(t)$ satisfies the following inequality:
\begin{eqnarray}
\label{lin_stab}
0 & >&\frac{\chi_{\infty} - \overline{\chi}(t)}{\overline{\chi}(t)^{2}} - 1 , \\
\mbox{which defines} & & \\
\overline{\chi}(t) & > & \frac{1}{2} \left( -1 + \sqrt{1 + 4 \chi_{\infty}} \right) \equiv \chi_{\mbox{crit}}.
\end{eqnarray}
This analysis indicates if the initial effective temperature is above a critical value,  $\chi_0 > \chi_{\mbox{crit}}$, then small amplitude perturbations are stable and the system will not localize.  However, if the initial effective temperature is less than this value (plus a diffusive term that depends on the wavenumber), there exists a period of time where small perturbations to the effective temperature will grow.  A time-varying stability analysis of thermal STZ equations by Foglia shows that the effective temperature is transiently unstable to perturbations in that context as well~\cite{Foglia}.

  This description of perturbations to a spatially homogeneous, time-varying, ``start-up'' trajectory is remarkably similar that given by Fielding and Olmsted~\cite{Fielding} in a study of shear banding in the Johnson-Segalman model for complex fluids. The stability of the solution to perturbations is simpler to analyze in the STZ formulation than in the complex fluid model because the linearized evolution of perturbations to $\chi$ are decoupled from the stress, as shown in Eq.~(\ref{s_dec}). This suggests that the mechanism for shear band nucleation in ``start-up'' flows may be similar in a variety of materials.

\subsection{Finite amplitude effects and a criterion for localization}
\label{finite}
  While linear stability analysis determines when perturbations grow, by itself it does not provide enough information to determine whether or not the integrated STZ equations exhibit strain localization. Many of the numerical STZ solutions for an unstable initial effective temperature do not localize if the initial perturbations are ``too small''.  For example, numerical integration of the STZ equations with an unstable initial value for the effective temperature $\chi_0 = 0.09 <  \chi_{\mbox{crit}} \simeq 0.13$ and a small initial perturbation $\delta \chi_{0} = 0.001$, result in a solution that is virtually indistinguishable from a homogeneous system. The purpose of this section is to derive a criterion based only on initial conditions (the mean value of the initial effective temperature $\chi_0$ and the amplitude of initial perturbations $\delta \chi_{0}$) that determines which materials will exhibit long-time strain localization.

 At first glance, the fact that a finite perturbation is required to generate localized solutions seems to contradict our assertion that the system is linearly unstable. However, linear stability equations are only valid for short times when perturbations are small -- they give no information about the long-time behavior of the unstable perturbed/inhomogeneous states. Nonlinear system dynamics can enhance or inhibit localization in the inhomogeneous states.  Upon further examination, we recognize that the ``finite-amplitude effect'' is due to nonlinear dynamics involving the interaction of two dynamical processes: perturbation growth and energy dissipation. Based on this understanding we derive a criterion for which initial conditions result in shear banding.

  We first show that rate of energy dissipation as a function of position determines whether perturbed states remain inhomogeneous or become uniform. Our guiding intuition is that perturbations that grow quickly permit most of the energy to be dissipated in the incipient shear band, which enhances localization. Otherwise the energy is dissipated throughout the material, which inhibits localization.
 
 Eq.~(\ref{simple_chi3}) can be rewritten in terms of the rate of plastic work dissipated by the system, $Q$:
\beq
\dot{s} = 1 - \frac{\overline{Q}}{s},
\eeq 
 where $\overline{Q}$ is spatial average of the rate of plastic work $Q(y)$. $Q$ can be resolved into a product of two terms -- the rate of plastic work per STZ, $\Gamma(s)$, that depends only on the stress and is spatially invariant, and the density of STZs, $\Lambda(\chi)$, which depends only on the effective temperature and is a function of position:
\beq
Q = \frac{2 \epsilon_0}{q_0} s {\cal C}(s) (1 - m) e^{-1/ \chi} \equiv \Gamma(s) \Lambda(\chi).
\eeq
 In steady state the average $\overline{Q} = \Gamma(s)\overline{\Lambda}$ is constrained to be a constant, but the value $Q(y) = \Gamma(s) \exp \left[ -1/ \chi(y) \right]$ is not similarly constrained.  This admits the possibility that the plastic work dissipated is very small at some positions and large at others, and this {\em must} occur for the system to remain localized.  

  To see this, first note that if the effective temperature is below $\chi_\infty$, Eq.~(\ref{chi1}) requires that the effective temperature must always increase proportional to $Q (\chi_\infty - \chi)$ or diffuse --- there is no relaxation towards thermal temperature in this approximation. Therefore, a spatially heterogeneous effective temperature distribution will be {\em nearly stationary} only if $Q$ is very close to zero whereever $\chi \ne \chi_\infty$. In other words, strain localization will only persist if the rate of plastic work dissipated outside the shear band is very small.
 
  The rate at which plastic work is dissipated is proportional to $\exp \left[ -1/ \chi(y) \right]$. A rough assumption about the nonlinear, inhomogeneous effective temperature dynamics is that the largest initial perturbation of amplitude $\delta \chi_{0}$ continues to grow at the rate we derived using linear stability analysis: $\omega(t) \delta \chi(t)$, while the effective temperature in regions outside the incipient shear band grows at the rate of a homogeneous system, $ \dot{\overline{\chi}} = f(\overline{\chi}(t), s(t))$. This assumption is consistent with behavior seen in numerical solutions to the STZ equations. 

\begin{figure}[h]
\centering \includegraphics[height=6cm]{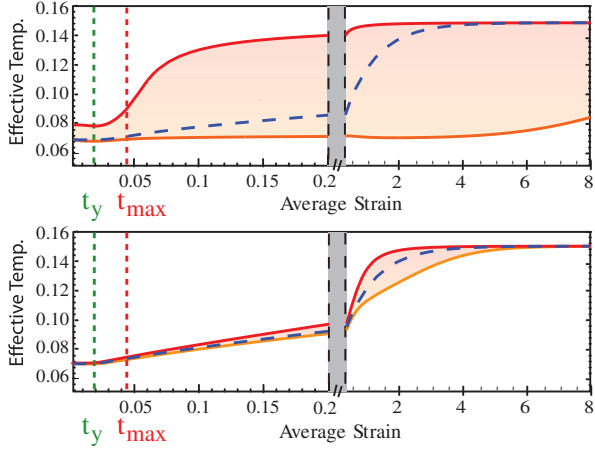}
\caption{\label{local_stab}(color online) Plot of effective temperature $\chi$ as a function of time for (a) an STZ solution for $\delta \chi_{0} = 0.01$ that localizes and (b) an STZ solution for $\delta \chi_{0} = 0.001$ that does not localize. The dashed blue line indicates the homogeneous solution for $\chi$ as a function of time. The red(orange) solid lines corresponds to the function $\chi(y)$ evaluated at a point located within (outside of) the shear band. The shaded region indicates the range of values spanned by $\chi(y)$ at every time, so that a large shaded region for long times indicates a long-lived localized state. Dashed vertical lines correspond to calculated values for the time the material first reaches the yield stress, $t_{y}$, and the time the material reaches its maximum stress $t_{max}$.}
\end{figure}

Figure~\ref{local_stab} shows the behavior of the effective temperature as a function of time for (a) a system that localizes and (b) a system that does not localize, illustrating the two different initial stages. In both plots the upper solid line is the maximum value of $\chi(y)$ inside the shear band, the lower solid line is the minimum value of $\chi(y)$ outside the shear band, and the dashed line is a homogeneous solution with the same mean value, $\overline{\chi}$. In plot (a), the perturbation grows quickly compared to the homogeneous solution for $\chi$, while in (b) the perturbation grows slowly. 

We are interested in whether the system dynamics focus energy dissipation inside the shear band or spread energy dissipation evenly throughout the material. Ideally, we would like to compare energy dissipation inside the band ($Q_{in}$) to energy dissipation outside of the band ($Q_{out}$) at various times $t$.  However, we only have information about the initial conditions, $\chi_0$ and $\delta \chi_{0}$.  Therefore we compare the time derivatives of $Q$ inside and outside the band: 
\bea
\label{Q_in_1}
\frac{\partial Q_{in}}{\partial t} &=& \Gamma(s) \frac{e^{-1/\chi_{in}}}{(\chi_{in})^{2}} \frac{\partial{\chi_{in}}}{\partial t} \nonumber \\
 &\simeq& \Gamma(s) \frac{e^{-1/\chi_{0}}}{(\chi_{0})^{2}} \: \omega_{\chi_0}(t) \: \delta \chi(t); \\
\frac{\partial Q_{out}}{\partial t} &=& \Gamma(s) \frac{e^{-1/\chi_{out}}}{(\chi_{out})^{2}} \frac{\partial{\chi_{out}}}{\partial t} \nonumber \\
\label{Q_out_1}
 &\simeq& \Gamma(s) \frac{e^{-1/\chi_{0}}}{(\chi_{0})^{2}} \; f(\overline{\chi}(t), s(t)).
\eea
We approximate $\chi_{in}$ and $\chi_{out}$ as the initial condition, $\chi_0$. This approximation is valid only for short times when $\chi$ has not changed appreciably from $\chi_0$ inside or outside the band. The ratio of the derivatives is given by: 
\beq
\label{Q_stab}
\frac{\dot{Q}_{in}}{\dot{Q}_{out}}(\delta \chi,\chi_0,t) = \frac{\omega_{\chi_0}(k, t) \delta \chi(t)}{f(\overline{\chi}(t), s(t))}.
\eeq
Equation~(\ref{Q_stab}) indicates that for short times when $\chi$ has not changed appreciably, the energy dissipation is determined by comparing the growth rate of perturbations to the growth rate of the mean effective temperature. 

Equation~(\ref{Q_stab}) must be evaluated at a particular time and wavenumber. We evaluate $\omega_{\chi_0}(k, t)$ for the most unstable mode, which corresponds to $k = \pi / L$. For notational simplicity this value of $k$ is assumed in the following derivation.  In writing Eqs.~(\ref{Q_in_1})~and~(\ref{Q_out_1}), we required that the time $t$ is small enough that $\chi$ remains close to $\chi_0$.  Numerical solutions confirm that $\overline{\chi}$ changes very little as $s$ increases from zero to its maximum shear stress. This stress, $s_{m}$, can be approximated as the solution to  Eq.~(\ref{def_g}) with $\dot{s} = 0$ and $\chi = \chi_0$: 
\beq
0 =  1 - 2 \frac{\epsilon_0}{q_0} {\cal C}(s_{m}) (1 - \frac{1}{s_{m}}) e^{-1 / \chi_0}
\eeq

 We must also approximate the numerator of Eq.~\ref{Q_stab} at the time when the perturbations are growing most rapidly. Naively, one might expect this to be the maximum of $\omega_{\chi_0}(t)$ times the amplitude of the initial perturbation, $\delta \chi_{0}$.  This underestimates the numerator because the amplitude of the perturbation increases with time.

 A better approximation can be found by noting that in numerical solutions, $\omega_{\chi_0}(t)$ rises almost linearly from zero at the time the material reaches the yield stress to a maximum at nearly the same time as the material attains its maximum stress, $s_{m}$. This behavior is seen in Fig.~\ref{a22}. Equation~(\ref{st_exp}) shows that $\omega_{\chi_0}(t)$ depends on $t$ only through $s(t)$ and $\overline{\chi}(t)$, so we evaluate the stability exponent at $(s_{m}, \chi_0)$.  Therefore we can approximate the linearized equation for the perturbations, Eq.~(\ref{chipert_dot}), as:
\beq
\dot{\delta \chi}(t) = \frac{\omega_{\chi_0}(s_{m},\chi_0)}{t_{m} - t_y} (t - t_y) \delta \chi(t) , \quad \quad t_{m}> t > t_y,
\eeq
where $t_y$ is the time when the material first reaches the yield stress and $t_{m}$ is the time at the maximum stress. The solution to this differential equation, evaluated at $s_{m}$, is
\beq
\delta \chi(t_{m}) = \delta \chi_{0} \; \exp[ \frac{\omega (s_{m}, \chi_0) \: (t_{m} - t_y)}{2} ].
\eeq
Therefore a lower bound on the numerator is given by:
\beq
\max_{t} \left[ \omega_{\chi_0}(t) \delta \chi(t)\right] > \omega(s_{m}, \chi_0)\: \delta \chi_{0} \:  \exp[ \frac{\omega(s_{m})\; \Delta t}{2} ],
\eeq
where $\Delta t = t_{m} - t_y$ is nearly constant in all the numerical STZ solutions and is about $0.03$ in units of strain. 

 While better than the first guess, this estimate is likely to be low because $\omega_{\chi_0}(t)$ is generally not sharply peaked about $t_{m}$, so the maximum growth rate occurs at later times than $t_{m}$. To account for this and to match the numerical solutions, we choose $\Delta t = 0.10$. 

Substituting these approximations into Eq.~\ref{Q_stab}, we define {\em localization ratio} ${\cal R}$ as follows:
\bea 
\label{R_stab}
{\cal R}(\delta \chi,\chi_0) &=&  \frac{ \omega (s_{m},\chi_0) \:  \delta \chi_{0} \: e^{\omega(s_{m},\chi_0) 0.05}} 
 { \mu^{*} \left( 1 - 2 \frac{\epsilon_0 {\cal C}(s_{m})}{q_0} (1 - \frac{1}{s_{m}})  e^{-1 / \chi_0} \right)},
\eea
where $\omega (s_{m},\chi_0)$ is defined by Eq.~(\ref{st_exp}). ${\cal R}$ is large if the system tends to focus energy dissipation inside the band and small otherwise. 

 We now systematically compare the value of the localization ratio to the degree of localization in a numerical STZ solution, and find that the ratio is an excellent criterion for localization. 

For each set of initial conditions, ($\chi_0$, $\delta \chi_0$), we compute a numerical solution to the STZ equations and then calculate the Gini coefficient for the strain rate as a function of position for each point in time. The Gini coefficient, $\phi(t)$, is a measure of localization and is defined as follows~\cite{Foglia_thesis,Gini}:
\bea
\phi(t) &=& \frac{1}{2 n^2 \overline{D^{pl}}} \sum_i \sum_j \vert {\cal D}^{pl}(y_i,t) -  {\cal D}^{pl}(y_j,t) \vert \nonumber \\
        &=&  \frac{1}{2 n^2 \overline{\Lambda}} \sum_i \sum_j \vert e^{-1/\chi(y_i,t)} -  e^{-1/\chi(y_j,t)} \vert ,
\eea
 where the function $\chi$ is evaluated at $n$ points in space, $\{y_n\}$. A discrete delta function has a Gini coefficient equal to unity, and a constant function has a Gini coefficient equal to zero. We define the {\em Numerical localization number} $\Phi$ as the maximum value of $\phi(t)$ over $t$. If $\Phi$ is close to unity the strain rate is highly localized (as in Figure~\ref{position_data}(b)) and when $\Phi$ is close to zero the strain rate remains homogeneous.

\begin{figure}[h]
\centering \includegraphics[height=7cm]{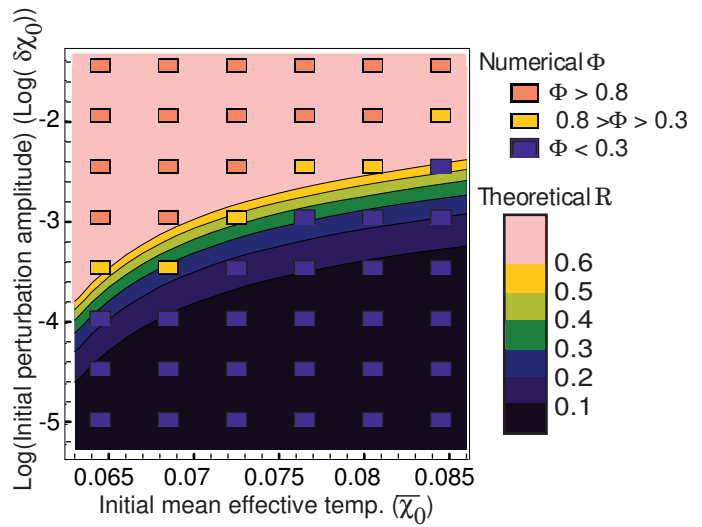}
\caption{\label{gini_loc}(color online) Comparison of numerical localization $\Phi$ to theoretical ${\cal R}$ as a function of $\chi_0$ and  $\log_{10}(\delta \chi_{0})$. Each rectangle in Figure~\ref{gini_loc} corresponds to a solution to the STZ equations for a particular choice of ($\chi_0$, $\delta \chi_{0}$). The shading of the rectangle corresponds to the value of $\Phi$ for a numerical solution with the indicated initial conditions: red corresponds to highly localized strain rate distributions, $1 > \Phi > 0.8$, yellow corresponds to partially localized solutions  $0.8 > \Phi > 0.3$ and dark blue corresponds to homogeneous solutions $0.3 > \Phi > 0$. The background shading corresponds to a contour plot for ${\cal R}$. Light pink corresponds to values of ${\cal R}> 0.6$, dark purple corresponds to values of ${\cal R}< 0.1$, and contours are at $0.1$ intervals in between. The yellow rectangles consistently overlay the yellow contour, showing that the theory and numerics agree on the localization transition.}
\end{figure}

Figure~\ref{gini_loc} shows that the numerical localization number $\Phi$ (rectangles) and the theoretical ratio ${\cal R}$ (contour plot) exhibit the same behavior as  a function of $\chi_0$ and  $\delta \chi_{0}$. This indicates that the theoretical ${\cal R}$ is a good criterion for the numerically computed strain localization $\Phi$. 

 The shading of the rectangles corresponds to the value of $\Phi$ for a given set of initial conditions: red corresponds to highly localized strain rate distributions ($\Phi > 0.8$),  yellow to partially localized solutions ($0.8 \geq \Phi \geq 0.3$), and dark blue to homogeneous solutions ($\Phi < 0.3$). The contour plot corresponds to values of ${\cal R}$: light pink for ${\cal R}> 0.6$, darker shades for ${\cal R} \leq 0.6$.  The data in figure~\ref{gini_loc} show that to a good approximation, ${\cal R}> 0.6$ corresponds to solutions with strain localization, and ${\cal R} \leq 0.6$ corresponds to homogeneous solutions.

 The localization ratio ${\cal R}$ is a function only of the STZ parameters and the initial conditions for the function $\chi(y)$. This is potentially a very useful tool for incorporating effects of localization into constitutive laws. It determines an important material property without requiring computation of the full inhomogeneous system dynamics.

\subsection{Long-time behavior}
\label{long_time}

 The localization ratio compares the energy dissipated inside the shear band to the energy dissipated outside the shear band, and determines whether the nonlinear dynamics enhance or inhibit shear band formation. An interesting and more difficult question is what determines whether the system dynamics result in a single shear band or multiple shear bands.  The MD simulations generally have a single shear band except at the highest strain rates (see Figure 2 in~\cite{Shi}). In the parameter range used to match the MD simulations, the STZ solutions also exhibit a single shear band.  However, outside this parameter range it is possible for the STZ equations to generate solutions with multiple shear bands.

 It appears that the STZ equations generate a single shear band when the system is in a parameter range where perturbations to $\chi$ are highly unstable.  In this case the largest amplitude perturbation grows rapidly, resulting in a large rate of energy dissipation at that position. Consequently the energy dissipation rates outside the shear band are smaller, which inhibits the growth of smaller amplitude perturbations. A full analysis of these dynamics is beyond the scope of this paper. 

  However, we use the fact that there is a single shear band to determine the width of that band. As discussed above, most of the strain must be accommodated in the shear band if the strain is to remain localized.  $\chi_{\infty}$ sets the steady state of disorder in the system, and we assume that $\chi$ attains this value in the shear band.  We can estimate the width of the band as a function of time, $w_E(t)$, by postulating that {\em all} of the shear band strain is accommodated plastically in the shear band:
\bea
0 &=& \overline{\dot{\gamma}} - \frac{1}{2 L} \int_{-L}^{L} D^{pl}(y) dy ; \\
0 &=& 1 - \frac{w_E}{2 L} \frac{2}{q_0}{\cal C}(s(t))\left( 1 - \frac{1}{s(t)} \right) e^{ -1/ \chi_{\infty}}; \\
\frac{w_E(t)}{2 L} &=& \frac{q_0 \exp[ 1/ \chi_{\infty}]}{2 {\cal C}(s(t))(1 - \frac{1}{s(t)})},
\eea
where $s(t)$ is the numerical solution to the STZ equation for the stress as a function of time.

\begin{figure}[h]
\centering \includegraphics[width=7cm]{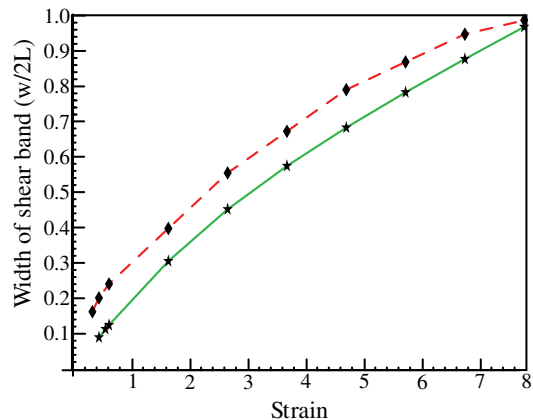}
\caption{\label{width}(color online) Data for shear band width as a function of strain. The red dashed line is the numerical width $w_N$ calculated from the STZ solution for the strain rate. The solid green line is the estimated shear band width $w_E$ calculated from the STZ solution for the stress only.}
\end{figure}

  We also evaluate the width of the shear band in the numerical STZ solution for the strain rate. The numerical shear band width ($w_N(t)$) is computed as the width of the strain rate vs. position curve at a value corresponding to the imposed strain rate, $1 = V_0 / L$. Figure~\ref{width} compares the numerical shear band width to the estimated width, as a function of time. The two are in agreement. The numerically computed width is systematically larger than the estimated width because we chose a relatively low cutoff, $1 = V_0 / L$, for the strain rate in the band.

 The estimated width is not a prediction for the width of the shear band because it depends on the numerical solution for the stress as a function of time.  However, it does suggest that because the STZ equations contain no natural length scale for the width of a shear band, the system dynamically sets the width based on the imposed strain rate.

 The shear band has a well-defined effective temperature ($\chi_{\infty}$) and therefore accommodates a fixed rate of plastic strain per unit width at a given stress. Together, the stress and the imposed strain set the shear band width: the shear band must grow to the width required (at a fixed stress) to accommodate all of the imposed strain.  While we cannot compute the shear band width analytically, we have recast the problem in terms of a potentially simpler one to solve: determining the stress relaxation as a function of time.

The value of the steady state stress has a large effect on strain localization.  While the spatial dependence of $Q(y)$ is through $\chi(y)$, the magnitude of $Q$ depends on the value of the stress through $\Gamma(s)$. When the system is in the flowing regime, the nondimensionalized stress $s$ is always greater than one.  The energy dissipation per STZ,  $\Gamma(s) \propto  s {\cal C}(s) (1 - 1/s)$, approaches zero as the stress approaches one from above. Therefore, if the steady state stress is very close to unity, the effective temperature dynamics become very slow -- they are ``frozen in'' by the stress dynamics.  In this parameter range the rate of plastic work can become very small at some positions as the system approaches steady state and this is exactly what permits a localized strain state to be long-lived. In contrast, when the steady state stress is too large the magnitude of $Q$ is also large (regardless of $\chi(y)$) and the material will not remain localized.
  
  Numerical integration of the STZ equations confirms that localization occurs only when the steady state stress is close to unity. STZ solutions exhibit localization when $q_0$ is ${\cal O} (10^{-6})$ and $\chi_{\infty}$ is ${\cal O}(1/10)$, which corresponds to a steady state stress that is very close to the yield stress ($s_f = 1.005$). However, they do not exhibit long-lived shear bands if the system is driven more quickly ($q_0 \sim {\cal O}(10^{-5})$), which corresponds to a steady state stress of $s_f \simeq 1.1$.

 This helps us interpret MD simulation data at higher strain rates.  In addition to the data shown in Figure~\ref{match_stress_falk}, Shi, {\em et al.} have computed stress-strain curves for two higher strain strain rates (five and 25 times greater, respectively, than the strain rate used to generate Fig.~\ref{match_stress_falk})~\cite{Shi2}. 

The lowest strain rate is denoted {\em V1} and the higher strain rates are denoted {\em V2} and {\em V3}. In this paper we defined the timescale ratio $q_0$ using {\em V1}. Therefore {\em V2} corresponds to $5 q_0$ and {\em V3} corresponds to  $25 q_0$. 

We numerically integrate the STZ equations at $5 q_0$ and $25 q_0$ and find that the resulting stress-strain curves do not match the simulation data. The theoretical STZ steady state flow stress is the same for all three strain rates, while in the MD simulations the flow stress for a system driven at {\em V3} is significantly ($\sim$ 15\%) higher.

One way to eliminate this discrepancy is to assume that we have misidentified the original timescale ratio $q_0$. For a given value of $q_0$ (and $\chi_{\infty}$), we can calculate the flow stress $s_f$ using Eq.~(\ref{sf}). To match the simulation data, the flow stress that corresponds to $25 q_0$ should be 15\% higher than the flow stress that corresponds to $q_0$. In Table~\ref{q0_table} we have calculated the flow stress ratios for several values of $q_0$.
\begin{table}
\begin{tabular}{|c|r|r|}
\hline
$q_0$ & $s_f( 5 q_0)/s_f( q_0) $  & $s_f(25 q_0)/s_f( q_0) $\\ \hline
$1.6 \times 10^{-6} $  & 1.005 & 1.028 \\ \hline
$1 \times 10^{-5} $ & 1.028 & 1.144 \\ \hline
$1 \times 10^{-4} $ & 1.185 & 1.695 \\ \hline
\end{tabular}
\caption{ \label{q0_table} Comparison of the STZ steady-state flow stress for different values of $q_0$.}
\end{table}
  This table shows that $q_0 \simeq 1 \times 10^{-5}$ better fits the MD simulation data for different strain rates than $q_0 \simeq 1 \times 10^{-6}$. The higher value for $q_0$ also makes sense physically because it corresponds to an STZ timescale $\tau_0$ which is simply the natural timescale for the Lennard-Jones glass, $t_0$.

 Our athermal model requires a small value for $q_0$ to generate solutions with long-lived localized states, but a larger value of $q_0$ to be consistent with MD simulations data for different strain rates. This disparity is likely an indication that the athermal approximation does not adequately capture long-time behavior of Lennard-Jones glasses.

 In an athermal description, the effective temperature always tends towards $\chi_{\infty}$ because there is no thermal relaxation.  Alternatively, if the effective temperature can relax towards a thermal bath temperature in regions where there is little plastic deformation, localized states can persist even when $s_f$ is large. It seems likely that a thermal description would generate localized solutions with higher values for $q_0$. Although the athermal STZ formulation reproduces many aspects of the MD simulations, a model with thermal relaxation may be required to capture more details of the MD simulations. This will be a direction of future research.  

\section{Concluding Remarks}
\label{Conclusions}
We have investigated the athermal STZ model with effective temperature for amorphous materials in a simple shear geometry. In contrast to other studies, the effective temperature field varies as a function of position and by numerically integrating the STZ equations we have shown that the resulting solutions can exhibit strain localization. The numerical STZ solutions match the stress-strain curves for MD simulations of Lennard-Jones glasses by Shi, {\em et al.} and exhibit strain rate and effective temperature fields that are consistent with those seen in simulations.

The continuum STZ model provides a framework for understanding how shear bands nucleate and persist for long times. We have shown that the model is unstable with respect to small perturbations of the effective temperature during the transient dynamics, though it is stable to perturbations in steady state.  Interestingly, shear localization does not always occur when the system is transiently unstable because shear bands can only form if the perturbations grow quickly compared to the homogeneous solution. This is a nonlinear effect which is driven by uneven energy dissipation in the system. Using rough approximations to these nonlinear dynamics we were able to derive a localization criterion that depends only on the initial conditions for the effective temperature. 

In the parameter range studied, STZ theory predicts that the sheared material attains a state with a single shear band.   The effective temperature field reaches its steady state value $\chi_{\infty}$ in the shear band and remains close to its initial value outside the band.  To a good approximation, all of the strain imposed at the boundaries is accommodated in the band, and we determine the width of the shear band using the STZ expression for the plastic strain rate when $\chi = \chi_{\infty}$. This analysis implies the shear band width is not determined by an inherent length scale in the material, but instead by a dynamical scale set by the imposed strain rate.

 Interesting questions remain concerning the interaction between shear bands and material boundaries.  We evaluate the STZ equations with simple, periodic boundary conditions on the effective temperature field. In our numerical solutions, the location of the shear band depends on the details of the initial fluctuations in $\chi$ and therefore seems arbitrary on macroscopic scales. This is similar to what is seen in Shi, {\em et al.} for simulations with periodic boundary conditions. However, in other MD simulations and in experiments on bulk metallic glasses the boundary conditions are non-trivial and most likely play an important role in nucleating shear bands.  For example, in simulations by Varnik et al.~\cite{Varnik} the system is driven at the boundary by a rough, rigid layer and the material consistently develops shear bands at the boundary. Additional simulations by Falk and Shi~\cite{Falk_Shi} show that shear bands are more likely to occur in the presence of surface defects.

\begin{figure}[h]
\centering \includegraphics[height=5cm]{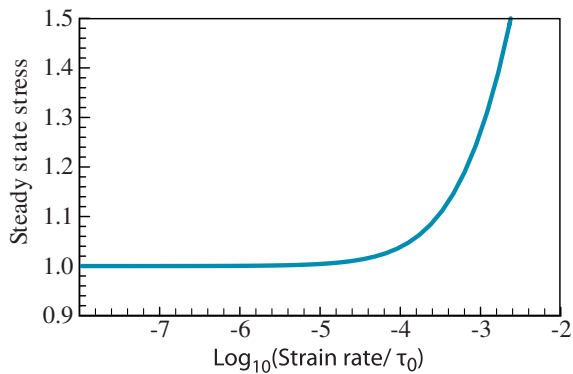}
\caption{\label{ssrate}STZ solution for the steady state stress vs. Log$_{10}$ of the imposed total strain rate. The curve is single-valued throughout, indicating that the system does not posses steady state coexisting regions with two different phases unless imposed by the boundary conditions.}
\end{figure}

  Boundary conditions may also influence the long-time behavior of shear bands. The long-lived, localized solutions to the STZ equations are not truly stationary states -- at long times the shear band will diffuse to cover the entire strip and the system will become homogeneous. Figure~\ref{ssrate} is a plot of the steady state stress vs. the natural log of the imposed strain-rate for STZ solutions. This function is not multivalued, indicating that the only steady state solution to the STZ equations with periodic boundary conditions is the spatially homogeneous solution $\chi(y) = \chi_\infty,\: s = s_f$.  Even excluding diffusion, the STZ equations will evolve to a homogeneous solution because the plastic dissipation rate is positive everywhere, which means the effective temperature is everywhere driven towards $\chi_\infty$, albeit at different rates.   However, if the effective temperature field is specified at the boundaries (i.e, the boundary causes ordering and fixes the disorder temperature there) then it is possible for solutions to have inhomogeneous steady states. This is similar to results for simulations with rough, rigid boundaries where the material develops {\em stationary} localized states~\cite{Varnik}.  More work is needed to understand the relationship between shear band development and material boundaries, and STZ theory should provide an excellent framework for these investigations.

The athermal STZ description provides a simple starting point for understanding the dynamics of shear band formation. As discussed in the previous section, a thermal description is likely a better model for the MD simulations at different strain rates.  Adding thermal relaxation to the STZ equations generates a picture of localization similar to those found in complex fluids, where the stress vs. strain rate curve is multi-valued~\cite{Lu_Olmsted}. As discussed by Langer~\cite{Langer_Eff} and Foglia~\cite{Foglia}, relaxation terms in the equation for the effective temperature generate solutions where the steady state stress is a multi-valued function of the strain-rate. In steady state, two  regions with different strain rates can coexist: one region which is nearly jammed except for thermal rearrangements and a second region where STZs are constantly created and flipped, resulting in plastic flow. 

This is different from the athermal theory where all of the material deforms due to shear-induced transitions, though the transitions occur at different rates. While athermal analysis provides a simple framework for understanding mechanisms leading to shear band nucleation, an STZ formulation with thermal relaxation may further enhance our understanding of steady-state dynamics and connect mechanisms for localization in amorphous solids to those in complex fluids. 

\section{Acknowledgments} The authors whould like to thank Yunfeng Shi and Michael Falk for numerous useful discussions and also for generously sharing their data from reference~\cite{Shi} and additional simulations~\cite{Shi2}. Their contributions were central to the analysis presented in this paper. M.L.M. would like to thank C. Maloney, A. Foglia, and G. Lois for useful discussions.  This work was supported by the James S. McDonnell Foundation, the David and Lucile Packard Foundation, and NSF grant number DMR-0606092. M.L.M. acknowledges an NSF Graduate Research Fellowship. J.S.L. was supported by DOE grant number DE-FG03-99ER45762.


\begin{thebibliography}{48}
\expandafter\ifx\csname natexlab\endcsname\relax\def\natexlab#1{#1}\fi
\expandafter\ifx\csname bibnamefont\endcsname\relax
  \def\bibnamefont#1{#1}\fi
\expandafter\ifx\csname bibfnamefont\endcsname\relax
  \def\bibfnamefont#1{#1}\fi
\expandafter\ifx\csname citenamefont\endcsname\relax
  \def\citenamefont#1{#1}\fi
\expandafter\ifx\csname url\endcsname\relax
  \def\url#1{\texttt{#1}}\fi
\expandafter\ifx\csname urlprefix\endcsname\relax\def\urlprefix{URL }\fi
\providecommand{\bibinfo}[2]{#2}
\providecommand{\eprint}[2][]{\url{#2}}

\bibitem[{\citenamefont{Shi et~al.}(2006)\citenamefont{Shi, Katz, Li, and
  Falk}}]{Shi}
\bibinfo{author}{\bibfnamefont{Y.}~\bibnamefont{Shi}},
  \bibinfo{author}{\bibfnamefont{M.~B.} \bibnamefont{Katz}},
  \bibinfo{author}{\bibfnamefont{H.}~\bibnamefont{Li}}, \bibnamefont{and}
  \bibinfo{author}{\bibfnamefont{M.}~\bibnamefont{Falk}},
  \bibinfo{journal}{Phys. Rev. Lett.} \textbf{\bibinfo{volume}{98}}
   \bibinfo{pages}{185505}(\bibinfo{year}{2007}).

\bibitem[{\citenamefont{Anandarajah et~al.}(2002)\citenamefont{Anandarajah,
  Sobhan, and Kuganenthira}}]{anandar}
\bibinfo{author}{\bibfnamefont{A.}~\bibnamefont{Anandarajah}},
  \bibinfo{author}{\bibfnamefont{K.}~\bibnamefont{Sobhan}}, \bibnamefont{and}
  \bibinfo{author}{\bibfnamefont{N.}~\bibnamefont{Kuganenthira}},
  \bibinfo{journal}{J. of Geotechnical Engineering}
  \textbf{\bibinfo{volume}{121}}, \bibinfo{pages}{57} (\bibinfo{year}{2002}).

\bibitem[{\citenamefont{Lauridsen et~al.}(2002)\citenamefont{Lauridsen,
  Twardos, and Dennin}}]{Lauridsen}
\bibinfo{author}{\bibfnamefont{J.}~\bibnamefont{Lauridsen}},
  \bibinfo{author}{\bibfnamefont{M.}~\bibnamefont{Twardos}}, \bibnamefont{and}
  \bibinfo{author}{\bibfnamefont{M.}~\bibnamefont{Dennin}},
  \bibinfo{journal}{Phys. Rev. Lett.} \textbf{\bibinfo{volume}{89}},
  \bibinfo{pages}{098303} (\bibinfo{year}{2002}).

\bibitem[{\citenamefont{Lu et~al.}(2003)\citenamefont{Lu, Ravichandran, and
  Johnson}}]{Lu}
\bibinfo{author}{\bibfnamefont{J.}~\bibnamefont{Lu}},
  \bibinfo{author}{\bibfnamefont{G.}~\bibnamefont{Ravichandran}},
  \bibnamefont{and} \bibinfo{author}{\bibfnamefont{W.}~\bibnamefont{Johnson}},
  \bibinfo{journal}{Acta Mater.} \textbf{\bibinfo{volume}{51}},
  \bibinfo{pages}{3429} (\bibinfo{year}{2003}).

\bibitem[{\citenamefont{Fenistein and Van~Hecke}(2003)}]{fenistein}
\bibinfo{author}{\bibfnamefont{D.}~\bibnamefont{Fenistein}} \bibnamefont{and}
  \bibinfo{author}{\bibfnamefont{M.}~\bibnamefont{Van~Hecke}},
  \bibinfo{journal}{Nature} \textbf{\bibinfo{volume}{425}},
  \bibinfo{pages}{256} (\bibinfo{year}{2003}).

\bibitem[{\citenamefont{Tsai and Gollub}(2005)}]{Tsai_Gollub}
\bibinfo{author}{\bibfnamefont{J.-C.} \bibnamefont{Tsai}} \bibnamefont{and}
  \bibinfo{author}{\bibfnamefont{J.~P.} \bibnamefont{Gollub}},
  \bibinfo{journal}{Phys. Rev. E} \textbf{\bibinfo{volume}{72}},
  \bibinfo{eid}{051304} (pages~\bibinfo{numpages}{10}) (\bibinfo{year}{2005}).

\bibitem[{\citenamefont{Kabla and Debr{\'e}geas}(2003)}]{Kabla}
\bibinfo{author}{\bibfnamefont{A.}~\bibnamefont{Kabla}} \bibnamefont{and}
  \bibinfo{author}{\bibfnamefont{G.}~\bibnamefont{Debr{\'e}geas}},
  \bibinfo{journal}{Phys. Rev. Lett.} \textbf{\bibinfo{volume}{90}},
  \bibinfo{pages}{258303} (\bibinfo{year}{2003}).

\bibitem[{\citenamefont{Mair and Callaghan}(1996)}]{mair}
\bibinfo{author}{\bibfnamefont{R.}~\bibnamefont{Mair}} \bibnamefont{and}
  \bibinfo{author}{\bibfnamefont{P.}~\bibnamefont{Callaghan}},
  \bibinfo{journal}{Europhys. Lett.} \textbf{\bibinfo{volume}{36}},
  \bibinfo{pages}{719} (\bibinfo{year}{1996}).

\bibitem[{\citenamefont{Makhloufi et~al.}(1995)\citenamefont{Makhloufi,
  Decruppe, Ait-Ali, and Cressely}}]{micelle2}
\bibinfo{author}{\bibfnamefont{R.}~\bibnamefont{Makhloufi}},
  \bibinfo{author}{\bibfnamefont{J.}~\bibnamefont{Decruppe}},
  \bibinfo{author}{\bibfnamefont{A.}~\bibnamefont{Ait-Ali}}, \bibnamefont{and}
  \bibinfo{author}{\bibfnamefont{R.}~\bibnamefont{Cressely}},
  \bibinfo{journal}{Europhys. Lett.} \textbf{\bibinfo{volume}{32}},
  \bibinfo{pages}{253} (\bibinfo{year}{1995}).

\bibitem[{\citenamefont{Johnson et~al.}(2002)\citenamefont{Johnson, Lu, and
  Demetriou}}]{Johnson2}
\bibinfo{author}{\bibfnamefont{W.}~\bibnamefont{Johnson}},
  \bibinfo{author}{\bibfnamefont{J.}~\bibnamefont{Lu}}, \bibnamefont{and}
  \bibinfo{author}{\bibfnamefont{M.}~\bibnamefont{Demetriou}},
  \bibinfo{journal}{Intermetallics} \textbf{\bibinfo{volume}{10}},
  \bibinfo{pages}{1039} (\bibinfo{year}{2002}).

\bibitem[{\citenamefont{Marone et~al.}(1990)\citenamefont{Marone, Raleigh, and
  Scholz}}]{Marone}
\bibinfo{author}{\bibfnamefont{C.}~\bibnamefont{Marone}},
  \bibinfo{author}{\bibfnamefont{C.}~\bibnamefont{Raleigh}}, \bibnamefont{and}
  \bibinfo{author}{\bibfnamefont{C.}~\bibnamefont{Scholz}},
  \bibinfo{journal}{J. Geophys. Res} \textbf{\bibinfo{volume}{95}},
  \bibinfo{pages}{7007} (\bibinfo{year}{1990}).

\bibitem[{\citenamefont{Spaepen}(1977)}]{Spaepen}
\bibinfo{author}{\bibfnamefont{F.}~\bibnamefont{Spaepen}},
  \bibinfo{journal}{Acta Metall.} \textbf{\bibinfo{volume}{25}},
  \bibinfo{pages}{407} (\bibinfo{year}{1977}).

\bibitem[{\citenamefont{Argon}(1979)}]{Argon}
\bibinfo{author}{\bibfnamefont{A.}~\bibnamefont{Argon}}, \bibinfo{journal}{Acta
  Metall.} \textbf{\bibinfo{volume}{27}}, \bibinfo{pages}{47}
  (\bibinfo{year}{1979}).

\bibitem[{\citenamefont{Bulatov and Argon}(1994)}]{Bulatov}
\bibinfo{author}{\bibfnamefont{V.~V.} \bibnamefont{Bulatov}} \bibnamefont{and}
  \bibinfo{author}{\bibfnamefont{A.~S.} \bibnamefont{Argon}},
  \bibinfo{journal}{Model. Simul. Mater. Sci. Eng.}
  \textbf{\bibinfo{volume}{2}}, \bibinfo{pages}{1994} (\bibinfo{year}{1994}).

\bibitem[{\citenamefont{Falk and Langer}(1998)}]{Falk_L1}
\bibinfo{author}{\bibfnamefont{M.}~\bibnamefont{Falk}} \bibnamefont{and}
  \bibinfo{author}{\bibfnamefont{J.}~\bibnamefont{Langer}},
  \bibinfo{journal}{Phys. Rev. E} \textbf{\bibinfo{volume}{57}},
  \bibinfo{pages}{7192} (\bibinfo{year}{1998}).

\bibitem[{\citenamefont{Bouchbinder et~al.}(2006)\citenamefont{Bouchbinder,
  Langer, and Procaccia}}]{Bouchbinder}
\bibinfo{author}{\bibfnamefont{E.}~\bibnamefont{Bouchbinder}},
  \bibinfo{author}{\bibfnamefont{J.}~\bibnamefont{Langer}}, \bibnamefont{and}
  \bibinfo{author}{\bibfnamefont{I.}~\bibnamefont{Procaccia}},
  \bibinfo{journal}{arXiv} \textbf{\bibinfo{volume}{cond-mat/0611026}}
  (\bibinfo{year}{2006}).

\bibitem[{\citenamefont{Falk and Langer}(2000)}]{Falk_L2}
\bibinfo{author}{\bibfnamefont{M.}~\bibnamefont{Falk}} \bibnamefont{and}
  \bibinfo{author}{\bibfnamefont{J.}~\bibnamefont{Langer}},
  \bibinfo{journal}{MRS Bull.} \textbf{\bibinfo{volume}{25}},
  \bibinfo{pages}{40} (\bibinfo{year}{2000}).

\bibitem[{\citenamefont{Pechenik}(2005)}]{Pechenik}
\bibinfo{author}{\bibfnamefont{L.}~\bibnamefont{Pechenik}},
  \bibinfo{journal}{Phys. Rev. E} \textbf{\bibinfo{volume}{72}},
  \bibinfo{pages}{021507} (\bibinfo{year}{2005}).

\bibitem[{\citenamefont{Varnik et~al.}(2003)\citenamefont{Varnik, Bocquet,
  Barrat, and Berthier}}]{Varnik}
\bibinfo{author}{\bibfnamefont{F.}~\bibnamefont{Varnik}},
  \bibinfo{author}{\bibfnamefont{L.}~\bibnamefont{Bocquet}},
  \bibinfo{author}{\bibfnamefont{J.-L.} \bibnamefont{Barrat}},
  \bibnamefont{and} \bibinfo{author}{\bibfnamefont{L.}~\bibnamefont{Berthier}},
  \bibinfo{journal}{Phys. Rev. Lett.} \textbf{\bibinfo{volume}{90}},
  \bibinfo{pages}{095702} (\bibinfo{year}{2003}).

\bibitem[{\citenamefont{Xu et~al.}(2005)\citenamefont{Xu, O'Hern, and
  Kondic}}]{Xu_OHern2}
\bibinfo{author}{\bibfnamefont{N.}~\bibnamefont{Xu}},
  \bibinfo{author}{\bibfnamefont{C.~S.} \bibnamefont{O'Hern}},
  \bibnamefont{and} \bibinfo{author}{\bibfnamefont{L.}~\bibnamefont{Kondic}},
  \bibinfo{journal}{Phys. Rev. E} \textbf{\bibinfo{volume}{72}},
  \bibinfo{eid}{041504} (pages~\bibinfo{numpages}{10}) (\bibinfo{year}{2005}).

\bibitem[{\citenamefont{Dieterich}(1978)}]{Dieterich}
\bibinfo{author}{\bibfnamefont{J.}~\bibnamefont{Dieterich}},
  \bibinfo{journal}{Pure and Applied Geophys.} \textbf{\bibinfo{volume}{116}},
  \bibinfo{pages}{790} (\bibinfo{year}{1978}).

\bibitem[{\citenamefont{Ruina}(1983)}]{Ruina}
\bibinfo{author}{\bibfnamefont{A.}~\bibnamefont{Ruina}}, \bibinfo{journal}{J.
  of Geophys. Res.} \textbf{\bibinfo{volume}{88}}, \bibinfo{pages}{10359}
  (\bibinfo{year}{1983}).

\bibitem[{\citenamefont{Bazant}(2006)}]{Bazant}
\bibinfo{author}{\bibfnamefont{M.}~\bibnamefont{Bazant}},
  \bibinfo{journal}{Mech. of Mater.} \textbf{\bibinfo{volume}{38}},
  \bibinfo{pages}{717} (\bibinfo{year}{2006}).

\bibitem[{\citenamefont{Sollich et~al.}(1997)\citenamefont{Sollich, Lequeux,
  H{\'e}braud, and Cates}}]{Sollich}
\bibinfo{author}{\bibfnamefont{P.}~\bibnamefont{Sollich}},
  \bibinfo{author}{\bibfnamefont{F.}~\bibnamefont{Lequeux}},
  \bibinfo{author}{\bibfnamefont{P.}~\bibnamefont{H{\'e}braud}},
  \bibnamefont{and} \bibinfo{author}{\bibfnamefont{M.}~\bibnamefont{Cates}},
  \bibinfo{journal}{Phys. Rev. Lett.} \textbf{\bibinfo{volume}{78}},
  \bibinfo{pages}{2020} (\bibinfo{year}{1997}).

\bibitem[{\citenamefont{Langer}(2004)}]{Langer_Eff}
\bibinfo{author}{\bibfnamefont{J.}~\bibnamefont{Langer}},
  \bibinfo{journal}{Phys. Rev. E} \textbf{\bibinfo{volume}{70}},
  \bibinfo{pages}{041502} (\bibinfo{year}{2004}).

\bibitem[{\citenamefont{Mehta and Edwards}(1989)}]{Edwards}
\bibinfo{author}{\bibfnamefont{A.}~\bibnamefont{Mehta}} \bibnamefont{and}
  \bibinfo{author}{\bibfnamefont{S.}~\bibnamefont{Edwards}},
  \bibinfo{journal}{Physica A} \textbf{\bibinfo{volume}{157}},
  \bibinfo{pages}{1091} (\bibinfo{year}{1989}).

\bibitem[{\citenamefont{Cugliandolo et~al.}(1997)\citenamefont{Cugliandolo,
  Kurchan, and Peliti}}]{Cugliandolo}
\bibinfo{author}{\bibfnamefont{L.}~\bibnamefont{Cugliandolo}},
  \bibinfo{author}{\bibfnamefont{J.}~\bibnamefont{Kurchan}}, \bibnamefont{and}
  \bibinfo{author}{\bibfnamefont{L.}~\bibnamefont{Peliti}},
  \bibinfo{journal}{Phys. Rev. E} \textbf{\bibinfo{volume}{55}},
  \bibinfo{pages}{3898} (\bibinfo{year}{1997}).

\bibitem[{\citenamefont{Ono et~al.}(2002)\citenamefont{Ono, O'Hern, Durian,
  Langer, Liu, and Nagel}}]{Ono}
\bibinfo{author}{\bibfnamefont{I.~K.} \bibnamefont{Ono}},
  \bibinfo{author}{\bibfnamefont{C.~S.} \bibnamefont{O'Hern}},
  \bibinfo{author}{\bibfnamefont{D.}~\bibnamefont{Durian}},
  \bibinfo{author}{\bibfnamefont{S.~A.} \bibnamefont{Langer}},
  \bibinfo{author}{\bibfnamefont{A.~J.} \bibnamefont{Liu}}, \bibnamefont{and}
  \bibinfo{author}{\bibfnamefont{S.~R.} \bibnamefont{Nagel}},
  \bibinfo{journal}{Phys. Rev. Lett.} \textbf{\bibinfo{volume}{89}},
  \bibinfo{pages}{095703} (\bibinfo{year}{2002}).

\bibitem[{\citenamefont{O'~Hern et~al.}(2004)\citenamefont{O'~Hern, Liu, and
  Nagel}}]{OHern}
\bibinfo{author}{\bibfnamefont{C.}~\bibnamefont{O'~Hern}},
  \bibinfo{author}{\bibfnamefont{A.}~\bibnamefont{Liu}}, \bibnamefont{and}
  \bibinfo{author}{\bibfnamefont{S.}~\bibnamefont{Nagel}},
  \bibinfo{journal}{Phys. Rev. Lett.} \textbf{\bibinfo{volume}{93}},
  \bibinfo{pages}{165702} (\bibinfo{year}{2004}).

\bibitem[{\citenamefont{Griggs and Baker}(1969)}]{Griggs}
\bibinfo{author}{\bibfnamefont{D.}~\bibnamefont{Griggs}} \bibnamefont{and}
  \bibinfo{author}{\bibfnamefont{D.}~\bibnamefont{Baker}},
  \bibinfo{journal}{Properties of Matter Under Unusual Conditions.
  Wiley/Intersciences, New York} pp. \bibinfo{pages}{23--42}
  (\bibinfo{year}{1969}).

\bibitem[{\citenamefont{Lewandowski and Greer}(2005)}]{Lewandowski}
\bibinfo{author}{\bibfnamefont{J.}~\bibnamefont{Lewandowski}} \bibnamefont{and}
  \bibinfo{author}{\bibfnamefont{A.}~\bibnamefont{Greer}},
  \bibinfo{journal}{Nat. Mater.} \textbf{\bibinfo{volume}{5}},
  \bibinfo{pages}{15} (\bibinfo{year}{2005}).

\bibitem[{\citenamefont{Braeck and Podladchikov}(2006)}]{Braeck}
\bibinfo{author}{\bibfnamefont{S.}~\bibnamefont{Braeck}} \bibnamefont{and}
  \bibinfo{author}{\bibfnamefont{Y.}~\bibnamefont{Podladchikov}},
  \bibinfo{journal}{Phys. Rev. Lett.} \textbf{\bibinfo{volume}{98}},
  \bibinfo{pages}{095504} (\bibinfo{year}{2007}).

\bibitem[{\citenamefont{Lemaitre}(2002)}]{Anael1}
\bibinfo{author}{\bibfnamefont{A.}~\bibnamefont{Lemaitre}},
  \bibinfo{journal}{Phys. Rev. Lett.} \textbf{\bibinfo{volume}{89}},
  \bibinfo{pages}{195503} (\bibinfo{year}{2002}).

\bibitem[{\citenamefont{Lois et~al.}(2005)\citenamefont{Lois, Lema{\^\i}tre,
  and Carlson}}]{Gregg}
\bibinfo{author}{\bibfnamefont{G.}~\bibnamefont{Lois}},
  \bibinfo{author}{\bibfnamefont{A.}~\bibnamefont{Lema{\^\i}tre}},
  \bibnamefont{and} \bibinfo{author}{\bibfnamefont{J.}~\bibnamefont{Carlson}},
  \bibinfo{journal}{Phys. Rev. E} \textbf{\bibinfo{volume}{72}},
  \bibinfo{pages}{51303} (\bibinfo{year}{2005}).

\bibitem[{\citenamefont{Lifshitz and Pitaevskii}(1980)}]{Landau}
\bibinfo{author}{\bibnamefont{Lifshitz}} \bibnamefont{and}
  \bibinfo{author}{\bibnamefont{Pitaevskii}}, \emph{\bibinfo{title}{Statistical
  Physics 3rd Edition Part 1}} (\bibinfo{publisher}{Pergamon Press},
  \bibinfo{address}{Elmsford, N.Y.}, \bibinfo{year}{1980}).

\bibitem[{\citenamefont{Maloney and Lema{\^\i}tre}(2004)}]{Maloney}
\bibinfo{author}{\bibfnamefont{C.}~\bibnamefont{Maloney}} \bibnamefont{and}
  \bibinfo{author}{\bibfnamefont{A.}~\bibnamefont{Lema{\^\i}tre}},
  \bibinfo{journal}{Phys. Rev. Lett.} \textbf{\bibinfo{volume}{93}},
  \bibinfo{pages}{195501} (\bibinfo{year}{2004}).

\bibitem[{\citenamefont{Jop et~al.}(2006)\citenamefont{Jop, Forterre, and
  Pouliquen}}]{Jop}
\bibinfo{author}{\bibfnamefont{P.}~\bibnamefont{Jop}},
  \bibinfo{author}{\bibfnamefont{Y.}~\bibnamefont{Forterre}}, \bibnamefont{and}
  \bibinfo{author}{\bibfnamefont{O.}~\bibnamefont{Pouliquen}},
  \bibinfo{journal}{Nature} \textbf{\bibinfo{volume}{441}},
  \bibinfo{pages}{727} (\bibinfo{year}{2006}).

\bibitem[{\citenamefont{Urbakh et~al.}(2004)\citenamefont{Urbakh, Klafter,
  Gourdon, and Israelachvili}}]{urbakh}
\bibinfo{author}{\bibfnamefont{M.}~\bibnamefont{Urbakh}},
  \bibinfo{author}{\bibfnamefont{J.}~\bibnamefont{Klafter}},
  \bibinfo{author}{\bibfnamefont{D.}~\bibnamefont{Gourdon}}, \bibnamefont{and}
  \bibinfo{author}{\bibfnamefont{J.}~\bibnamefont{Israelachvili}},
  \bibinfo{journal}{Nature} \textbf{\bibinfo{volume}{430}},
  \bibinfo{pages}{525} (\bibinfo{year}{2004}).

\bibitem[{\citenamefont{Lema{\^\i}tre and Carlson}(2004)}]{Anael2}
\bibinfo{author}{\bibfnamefont{A.}~\bibnamefont{Lema{\^\i}tre}}
  \bibnamefont{and} \bibinfo{author}{\bibfnamefont{J.}~\bibnamefont{Carlson}},
  \bibinfo{journal}{Phys. Rev. E} \textbf{\bibinfo{volume}{69}},
  \bibinfo{pages}{61611} (\bibinfo{year}{2004}).

\bibitem[{\citenamefont{Daub and Carlson}(2007)}]{Daub}
\bibinfo{author}{\bibfnamefont{E.}~\bibnamefont{Daub}} \bibnamefont{and}
  \bibinfo{author}{\bibfnamefont{J.~M.} \bibnamefont{Carlson}},
  \bibinfo{journal}{submitted to J. Geo. Res.}  (\bibinfo{year}{2007}).

\bibitem[{\citenamefont{Foglia}(2006{\natexlab{a}})}]{Foglia}
\bibinfo{author}{\bibfnamefont{A.}~\bibnamefont{Foglia}},
  \bibinfo{journal}{arXiv} \textbf{\bibinfo{volume}{cond-mat/0608451}}
  (\bibinfo{year}{2006}{\natexlab{a}}).

\bibitem[{\citenamefont{Huang et~al.}(2002)\citenamefont{Huang, Suo, and
  Prevost}}]{Huang}
\bibinfo{author}{\bibfnamefont{R.}~\bibnamefont{Huang}},
  \bibinfo{author}{\bibfnamefont{Z.}~\bibnamefont{Suo}}, \bibnamefont{and}
  \bibinfo{author}{\bibfnamefont{W.}~\bibnamefont{Prevost},
  \bibfnamefont{J.H.and~Nix}}, \bibinfo{journal}{J. of the Mech. and Phys. of
  Solids} \textbf{\bibinfo{volume}{50}}, \bibinfo{pages}{1011}
  (\bibinfo{year}{2002}).

\bibitem[{\citenamefont{Fielding and Olmsted}(2003)}]{Fielding}
\bibinfo{author}{\bibfnamefont{S.~M.} \bibnamefont{Fielding}} \bibnamefont{and}
  \bibinfo{author}{\bibfnamefont{P.~D.} \bibnamefont{Olmsted}},
  \bibinfo{journal}{Phys. Rev. Lett.} \textbf{\bibinfo{volume}{90}},
  \bibinfo{pages}{224501} (\bibinfo{year}{2003}).

\bibitem[{\citenamefont{Foglia}(2006{\natexlab{b}})}]{Foglia_thesis}
\bibinfo{author}{\bibfnamefont{A.}~\bibnamefont{Foglia}}, Ph.D. thesis,
  \bibinfo{school}{University of California, Santa Barbara}
  (\bibinfo{year}{2006}{\natexlab{b}}).

\bibitem[{\citenamefont{Damgaard}(2007)}]{Gini}
\bibinfo{author}{\bibfnamefont{C.}~\bibnamefont{Damgaard}},
  \bibinfo{type}{Tech. Rep.}, \bibinfo{institution}{From MathWorld--A Wolfram
  Web Resource, created by Eric W. Weisstein} (\bibinfo{year}{2007}).

\bibitem[{\citenamefont{Shi and Falk}(2006)}]{Shi2}
\bibinfo{author}{\bibfnamefont{Y.}~\bibnamefont{Shi}} \bibnamefont{and}
  \bibinfo{author}{\bibfnamefont{M.}~\bibnamefont{Falk}},
  \bibinfo{howpublished}{private communication} (\bibinfo{year}{2006}).

\bibitem[{\citenamefont{Falk and Shi}(2003)}]{Falk_Shi}
\bibinfo{author}{\bibfnamefont{M.}~\bibnamefont{Falk}} \bibnamefont{and}
  \bibinfo{author}{\bibfnamefont{Y.}~\bibnamefont{Shi}}, \bibinfo{journal}{Mat.
  Res. Soc. Proc.} \textbf{\bibinfo{volume}{754}}, \bibinfo{pages}{20}
  (\bibinfo{year}{2003}).

\bibitem[{\citenamefont{Lu et~al.}(2000)\citenamefont{Lu, Olmsted, and
  Ball}}]{Lu_Olmsted}
\bibinfo{author}{\bibfnamefont{C.-Y.~D.} \bibnamefont{Lu}},
  \bibinfo{author}{\bibfnamefont{P.~D.} \bibnamefont{Olmsted}},
  \bibnamefont{and} \bibinfo{author}{\bibfnamefont{R.~C.} \bibnamefont{Ball}},
  \bibinfo{journal}{Phys. Rev. Lett.} \textbf{\bibinfo{volume}{84}},
  \bibinfo{pages}{642} (\bibinfo{year}{2000}).

\end{thebibliography}
\end{document}